\documentclass[12pt]{iopart}

\usepackage{iopams}  

\usepackage{bm}
\usepackage{amssymb}
\usepackage{graphicx}
\usepackage{float}
\usepackage{floatflt}

\begin{document}

\title{The Higgs Bridge}

\author{Roland E. Allen}
\address{Department of Physics and Astronomy \\ 
Texas A\&M University, College Station, Texas 77843, USA}

\begin{abstract}
The particle recently discovered at the Large Hadron Collider near Geneva is almost certainly a Higgs boson, the long-sought completion of the Standard Model of particle physics. But this discovery, an achievement by more than six thousand scientists (including students), is actually much more than a mere capstone of the Standard Model. It instead represents a bridge from the Standard Model to exciting discoveries of the future, at higher energies or in other experiments, and to the properties of matter at very low temperatures. The mere existence of a particle with zero spin implies a need for new physics, with the most likely candidate being supersymmetry, which requires that every known particle has a superpartner yet to be discovered. And phenomena similar to the Higgs are seen in superconducting metals and superfluid gases at low temperatures, which extend down to a millionth or even a billionth of a degree Kelvin. So the discovery of a Higgs boson has a central place in our attempts both to achieve a true understanding of Nature and to harness Nature in practical applications.
\end{abstract}

\maketitle

\section{\label{sec:sec1} Discovery of a new kind of particle}

\begin{figure}[htbp]
\centering
\includegraphics[bb=0 0 360 325, width=5in]{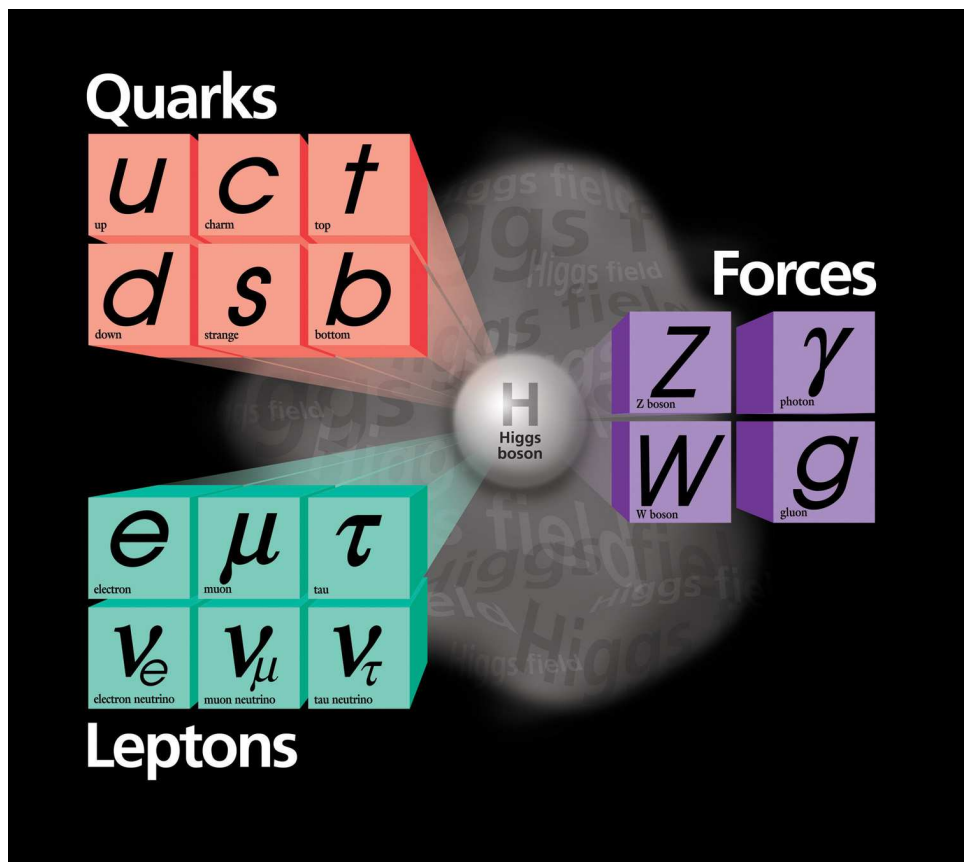}
\caption{The Higgs boson is the first fundamental particle to have no spin. It arises as an excitation of the Higgs field, which gives masses to  particles of the Standard Model: the \textit{fermions} on the left (six quarks, three electron-like leptons, and three neutrinos) and some of the  \textit{vector bosons} on the right: the $W$ and $Z$ particles, with the photon $\gamma$ and gluons $g$ remaining massless.  (The photon, the gluons, and the $W$ and $Z$ are respectively the carriers of the electromagnetic, strong nuclear, and weak nuclear forces.) The fermions all have spin 1/2, the vector bosons all have spin 1, and the  concept of spin is explained in the text. Credit: Fermilab Visual Media Services. \label{Fig1-FN0266H}}
\end{figure}

On July 4, 2012, after years, decades, and even generations of immense efforts by thousands of scientists, the world learned of the discovery of a new kind of particle: what is almost certainly a Higgs boson. Since the measured properties of this particle are now in excellent agreement with the required properties, in the remainder of this article we will drop the qualifier ``almost" and adopt the working assumption that this particle \textit{is} in fact a Higgs boson. Let us then proceed to describe why it deserves the world-wide exaltation with which it has been greeted, beginning with Fig.~\ref{Fig1-FN0266H}, which already provides some sense of the centrality of this particle. 

The Higgs boson is an excitation of the Higgs field, which permeates all of space. (As used here, the term ``Higgs field'' means only a single field in the context of the Standard Model, but includes at least a pair of such fields in the context of a supersymmetric extension.  One field is required to give masses to the top row of quarks in Fig.~\ref{Fig1-FN0266H}, and another for the bottom row.) As this field cooled with the rest of the universe, following the origin of the universe in an extremely hot Big Bang, it underwent a condensation, analogous to the condensation of water vapor into a liquid when it is cooled. 

\begin{figure}[htbp] 
\centering
 \includegraphics[bb=0 0 360 285, width=5in]{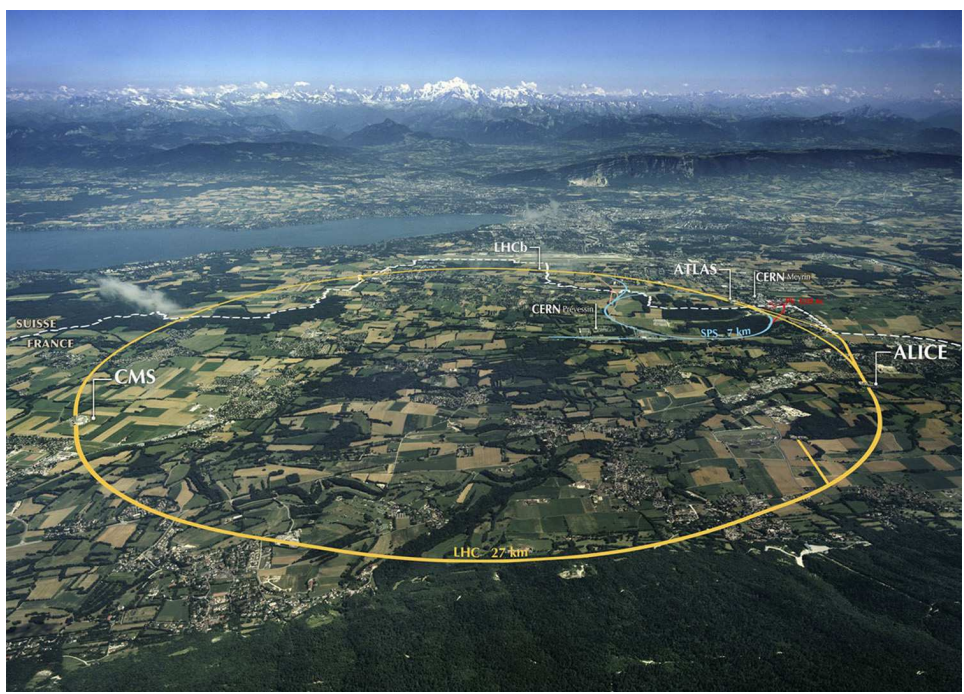}
 \caption{Aerial photograph representing the Large Hadron Collider, with the border between France and Switzerland indicated by a dashed line. The 27 km LHC tunnel cannot be seen, because it is 45 to 175 meters underground, but it is represented by the large circle. The four large detectors built around the collision points are discussed in Section \ref{sec:sec5}. Credit: CERN \label{Fig2-8701973.eps}}
 \end{figure}

But the Higgs condensate fills all of space, and you are always living and moving in this extremely dense fluid. You never notice it because it is essentially a superfluid with zero viscosity. To create an excitation of the Higgs condensate -- otherwise known as a Higgs boson -- requires smashing particles together at energies that could only be attained at the world's most powerful accelerator laboratories -- the Tevatron near Chicago and the Large Hadron Collider (or LHC) near Geneva. An aerial photo representing the LHC is shown in Fig.~\ref{Fig2-8701973.eps}.

These ideas have a rich history, which is briefly summarized elsewhere~\cite{Allen-EBH} and which centers on the \textit{Englert-Brout-Higgs mechanism}~\cite{London,Anderson,Englert,Higgs1,Higgs2,Kibble}. This mechanism explains how the $W$ and $Z$ particles of Fig.~\ref{Fig1-FN0266H} acquire their masses, and it played a leading role in the development of the Standard Model of particle physics~\cite{Weinberg,Salam}.

The masses of the $W^+$, $W^-$, and $Z$ are critically important because they account for the short range of the weak nuclear force, which causes radioactive beta decay and helps to power the Sun and other stars. According to the uncertainty principle of quantum mechanics, a force-carrying virtual particle can ``borrow'' an energy $\Delta E$ for a time $\Delta t$, where (in a simple picture) $\Delta E \Delta t \sim \hbar$ and $\hbar$ is Planck's constant divided by $2 \pi$. A particle with a large mass $M$ has to borrow  a large energy $Mc^2$, so it can only travel for a short time, and consequently a short distance. If there were no Higgs condensate to provide masses to the $W$ and $Z$ particles, the weak force would be long range, and we would not have the familiar slow nuclear burning of hydrogen in the Sun and other stars.

One interesting aspect of the Higgs discovery is that it demonstrates the unity of physics.  Phenomena in high energy physics, at enormous energies or temperatures, can be qualitatively the same as phenomena in condensed matter physics and atomic physics at quite low energies or temperatures. Examples in the present context are $^{4}$He, which begins to become a superfluid when cooled to 2.17 K, and various systems of trapped atoms, which have been cooled to less than $10^{-9}$ K. Superfluid $^{4}$He has exactly zero viscosity, so that an object moving through it effectively does not notice that it is there, until the object exceeds a critical velocity where it begins to produce excitations. In this sense, moving through $^{4}$He near 0 K would be like moving through the normally undetectable Higgs condensate.

An even more exciting aspect of this recent discovery is that it appears to be a harbinger of further revolutionary breakthroughs, as discussed below.

\section{\label{sec:sec2a} The Higgs as a bridge linking all of physics}

The Higgs is a bridge particle in a number of different respects. As implied by Fig.~\ref{Fig1-FN0266H}, it bridges the matter particles on the left and force particles on the right, providing masses to both kinds of particles, but in different ways. It also promises to be a bridge between the entire Standard Model (SM) of Fig.~\ref{Fig1-FN0266H} and new phenomena and concepts at higher energy: The mere existence of a scalar boson -- a particle with no spin -- requires new physics if its mass is to be protected from enormous quantum corrections, as explained below.

A particle's spin angular momentum $\mathbf{S}$ is analogous to the angular momentum of an ordinary spinning object, such as a bicycle wheel. But it is always the same for a given kind of particle, and it is quantized according to $S=\sqrt{s\left( s+1 \right)}\hbar$, where the spin quantum number $s$ is a multiple of $1/2$ -- i.e. $0$,  $1/2$, $1$, $3/2$, $\cdots$ .

All the matter and force particles of Fig.~\ref{Fig1-FN0266H} behave in essentially the same way as the electron and photon (particle of light) which are familiar in ordinary life. In fact they are so familar that, as you look around, all that you are observing is essentially just electrons and photons, with the atomic nuclei behaving as point charges for most purposes. The theory of electrons and photons was fully developed by the 1950s, and their behavior is well understood. In particular, it was found that quantum corrections lead to extremely weak (logarithmic) infinities that can be successfully controlled, in the sense that they do not appear in the results for physical quantities. 

As a consequence, the theoretical calculations and experimental measurements agree with an accuracy that is equivalent to measuring the distance between New York City and Los Angeles to within less than the thickness of a human hair. Calculations for the other matter and force particles at high energy use similar methods (which have been extended in various ways) and the infinities are still under control. These particles respectively have spin 1/2 and 1. 

But for the latest particle to enter the community of Fig.~\ref{Fig1-FN0266H} -- the Higgs boson with spin 0 -- the calculations lead to an extremely strong infinity for the correction to its mass $m_h$. If one imposes a cutoff energy $\Lambda_0$ in the calculation, then $m_h \rightarrow \infty$ in proportion to $\Lambda_0$, and some new physics is expected to intervene and cancel this infinite contribution to the particle mass.

\begin{figure}[htbp]
\centering
 \includegraphics[bb=0 0 360 165, width=5in]{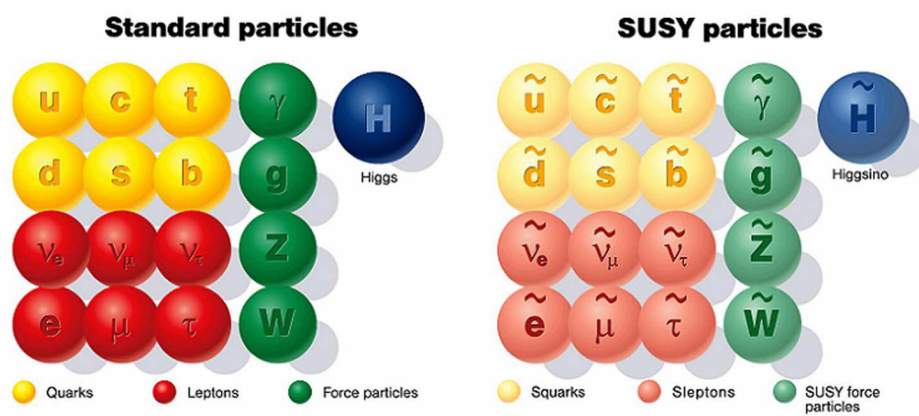}
 \caption{Particles of the Standard Model (SM) and their supersymmetric partners, called sparticles. The partners of spin 1/2 fermions are spin 0 sfermions. (A technical point: The right- and left-handed fields for each fermion, which are discussed below, actually have separate partners.) Since all the forces of the SM are described by \textit{gauge theories}, as described in the text, the spin 1 force particles are called gauge bosons, and their spin 1/2 partners are called gauginos. The partners of the spin 0 Higgs bosons are spin 1/2 Higgsinos. 
\newline \indent
There are four spin 1/2 partners with zero electric charge: two Higgsinos, a photino, and a zino, with the last corresponding to the $Z$. These can be combined to form four fields and particles with well-defined masses. One of these \textit{neutralinos} has the lowest mass, and is therefore stable in supersymmetic theories where (i) neutral sparticles have lower mass than charged sparticles and (ii) R-parity is conserved, with R-parity$=+1$ for ordinary particles and $-1$ for sparticles. 
\newline \indent
The lightest neutralino is therefore a candidate for the dark matter of the universe. Furthermore, since it experiences only the weak force (plus gravity), it turns out in detailed calculations to have emerged, from the hot universe following the Big Bang, with about the right density to explain the observed density of dark matter. Credit: DESY \label{Fig9-susyparticles_sm.eps}}
 \end{figure}

The most likely candidate is supersymmetry (susy), which provides supersymmetric partners whose effect exactly cancels that of the known particles. This beautiful cancellation is due to a minus sign difference in the contributions of the fermions, or spin 1/2 particles, and the bosons, or spin 1 and spin 0 particles. If the LHC does in fact discover susy after it resumes operations (in 2015), with its energy almost doubled from 8 to 14 TeV, it will mean both a new fundamental symmetry of Nature and a plethora of new \textit{sparticles} to match the known particles, as shown in Fig.~\ref{Fig9-susyparticles_sm.eps}.

The Higgs is also of central importance for other reasons. For example, it is related to the quest for a grand unified theory (GUT) of all the nongravitational forces:  the electromagnetic force, the strong nuclear force which holds quarks together in a proton or neutron (and protons and neutrons together in an atomic nucleus), and the weak nuclear force which causes radioactive beta decay. Such a theory will surely entail Higgs-like fields which condense at much higher energies. 

Finally, as discussed below, some of the greatest problems in cosmology are closely associated with the Higgs.  One is the \textit{cosmological constant problem} -- the fact that there should be a vacuum energy which is 50 or even 120 orders of magnitude larger than permitted by observation. People started worrying seriously about this problem after it was recognized that the vacuum should be occupied by the very dense Higgs condensate. 

Now there is also a second cosmological constant problem: Observations of type 1a supernovas (exploding white dwarfs), plus other astronomical observations, demonstrate that there is a \textit{dark energy} which behaves like a cosmological constant (or vacuum energy) -- relatively weak compared to the one of theory, but strong enough to dominate all other sources of gravity on a cosmic scale. 

\begin{figure}[htbp]
\centering
 \includegraphics[bb=0 0 360 260, width=5in]{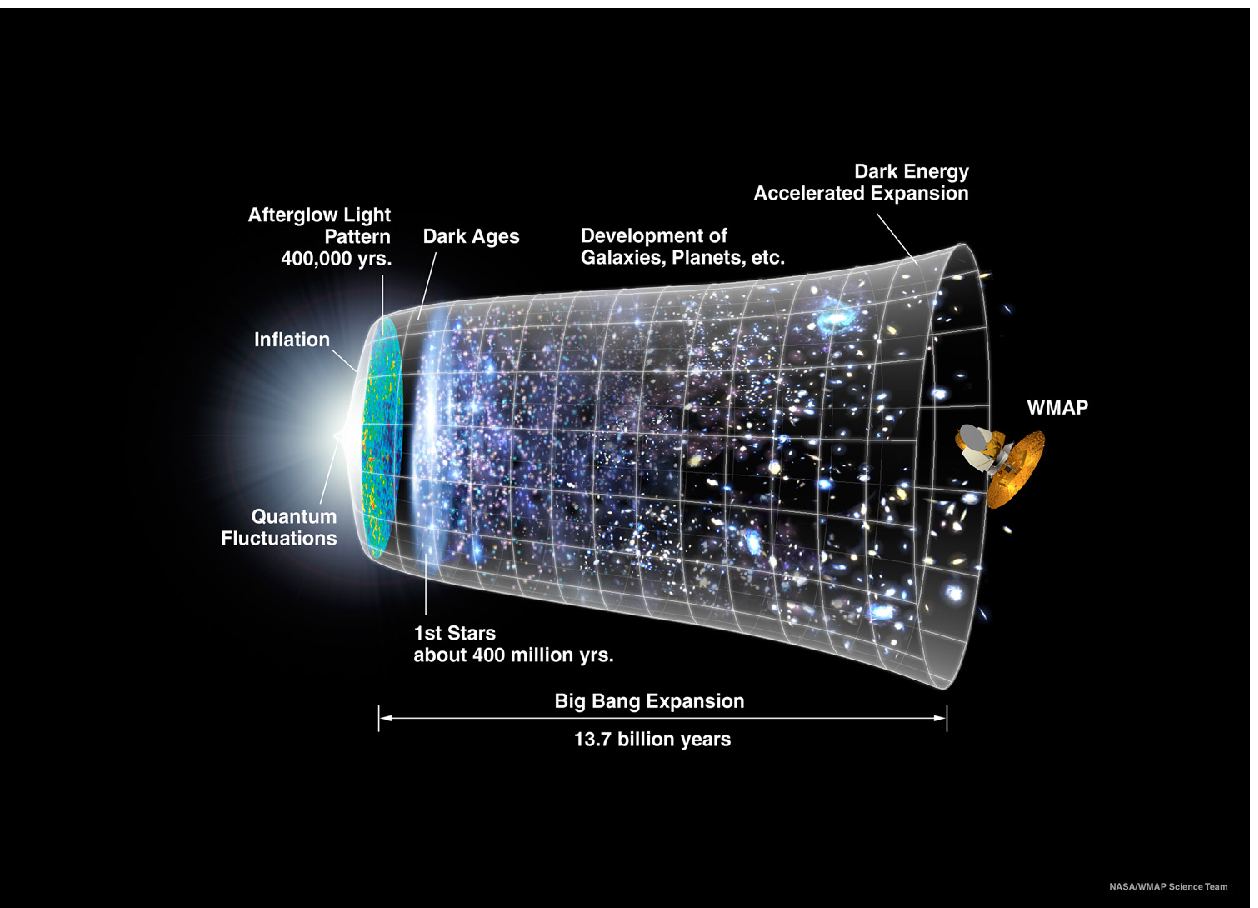}
 \caption{History of the universe, as inferred from the observations of the Planck, WMAP, and COBE satellites, confirmed and supplemented by many other astronomical studies. Credit: NASA / WMAP Science Team \label{Fig2b-CMB_Timeline.eps}}
 \end{figure}

\begin{figure}[htbp]
\centering
 \includegraphics[bb=0 0 360 180, width=5in]{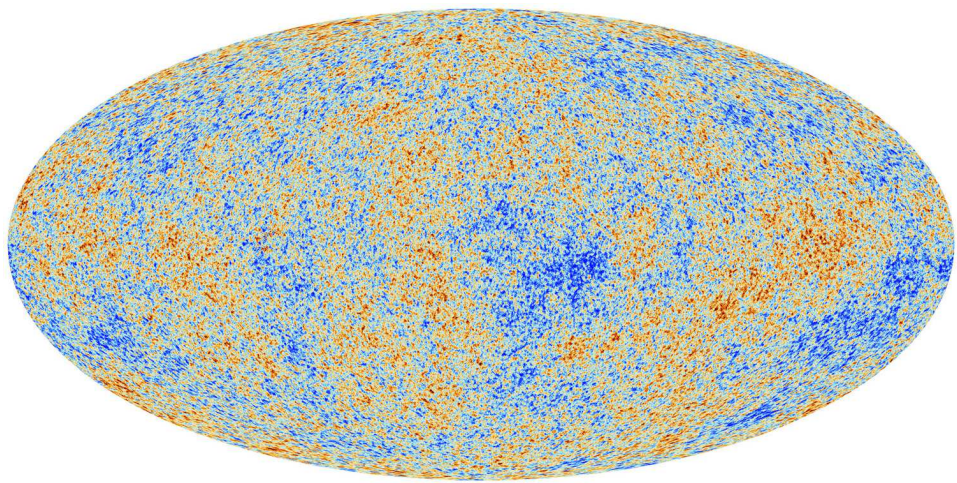}
 \caption{Temperature fluctuations, at the level of about one part in 100,000, observed by the Planck satellite. These observations provide a great deal of information about the early universe, including evidence for an extremely early \textit{inflationary epoch}, during which the universe is thought to have undergone an enormous expansion, as pictured in Fig.~\ref{Fig2b-CMB_Timeline.eps}. Credit: European Space Agency and the Planck Collaboration \label{Fig2a-planck-cmb.eps}}
 \end{figure}

Another problem related to the Higgs is the origin of \textit{inflation} in the early universe, as depicted in  Fig.~\ref{Fig2b-CMB_Timeline.eps}. Such an extremely rapid expansion of the universe (at about $10^{-35}$ second after its birth) would explain the nearly \textit{scale-free fluctuations} in the cosmic microwave background radiation. These have been observed by the Planck, WMAP, and COBE satellite experiments, and the results from Planck are shown in Fig.~\ref{Fig2a-planck-cmb.eps}. There is still no fully convincing model for inflation, but it would follow from the condensation of an appropriate ``inflaton'' field, quite similar to condensation of the Higgs field.

Below we will discuss in detail how the Higgs gives masses to the particles of Fig. 1. But it is worth mentioning why this is important. When one uses the Bohr model to calculate the radius of the orbit of an electron in its ground state in the hydrogen atom, the result is 
\begin{eqnarray}
r_1 = \frac{\hbar^2}{m_e k e^2} 
\label{eq1.1}
\end{eqnarray}
where $m_e$ is the mass of an electron, $k$ is the Coulomb's law constant, and $e$ is the fundamental electric charge. A full quantum calculation yields a probability distribution for the electron, rather than an orbit,  but gives the same result for the position of the peak in the radial probability density. As $m_e \rightarrow 0$,  the size of the atom then $\rightarrow \infty$. Similar results hold for other atoms, so without the Higgs condensate there would be no atoms or ordinary matter.

There are two basic origins of mass. The first was discovered by Albert Einstein and published in one of his miraculous 1905 papers: $E=mc^2$, or $m=E/c^2$.  (His other papers introduced the theory of relativity, the particle-wave duality of quantum mechanics, and convincing proof for atoms -- at the same time that the eminent physicist Ernst Mach was expressing strong opposition to the atomic theory). About 99\% of the mass of a proton, neutron, atom, or human body has this origin, in the kinetic energy of partons -- i.e., quarks and gluons -- whizzing around inside each nucleon. 

The Higgs mechanism discussed below represents the second way in which mass can originate: through the coupling of a particle to a bosonic field that undergoes condensation. 

\section{\label{sec:sec2} The known fundamental particles get their masses from the Higgs field}

\begin{figure}[htbp]
\centering
 \includegraphics[bb=0 0 360 240, width=5in]{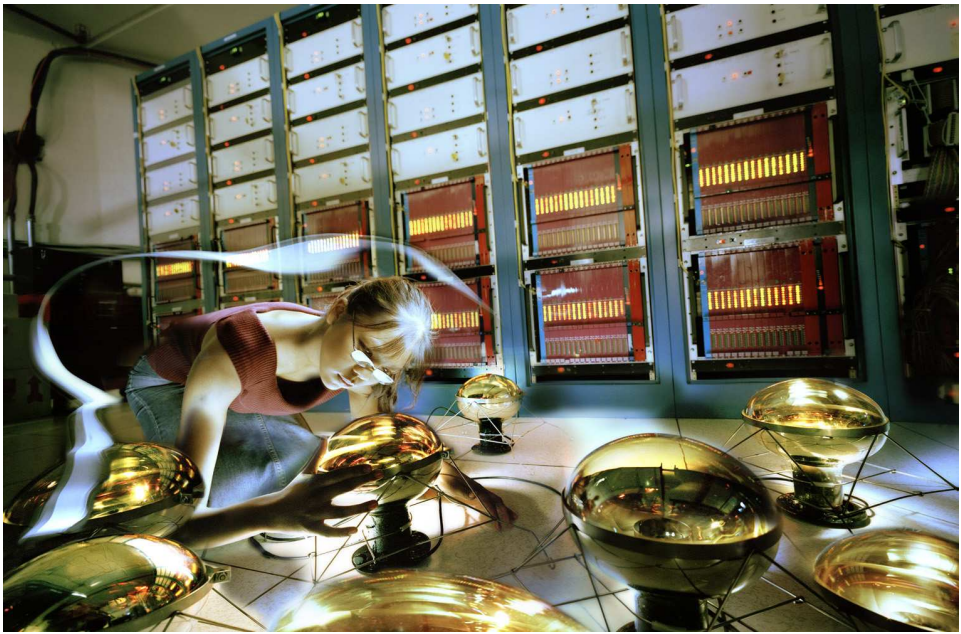}
 \caption{Jasmine Ma, a summer student working on the MiniBooNE experiment, inspects one of the phototubes that detect light from neutrino interactions. Credit: Peter Ginter \label{Fig3-FN0197H.eps}}
 \end{figure}

\begin{figure}[htbp]
\centering
 \includegraphics[bb=0 0 360 530, width=5in]{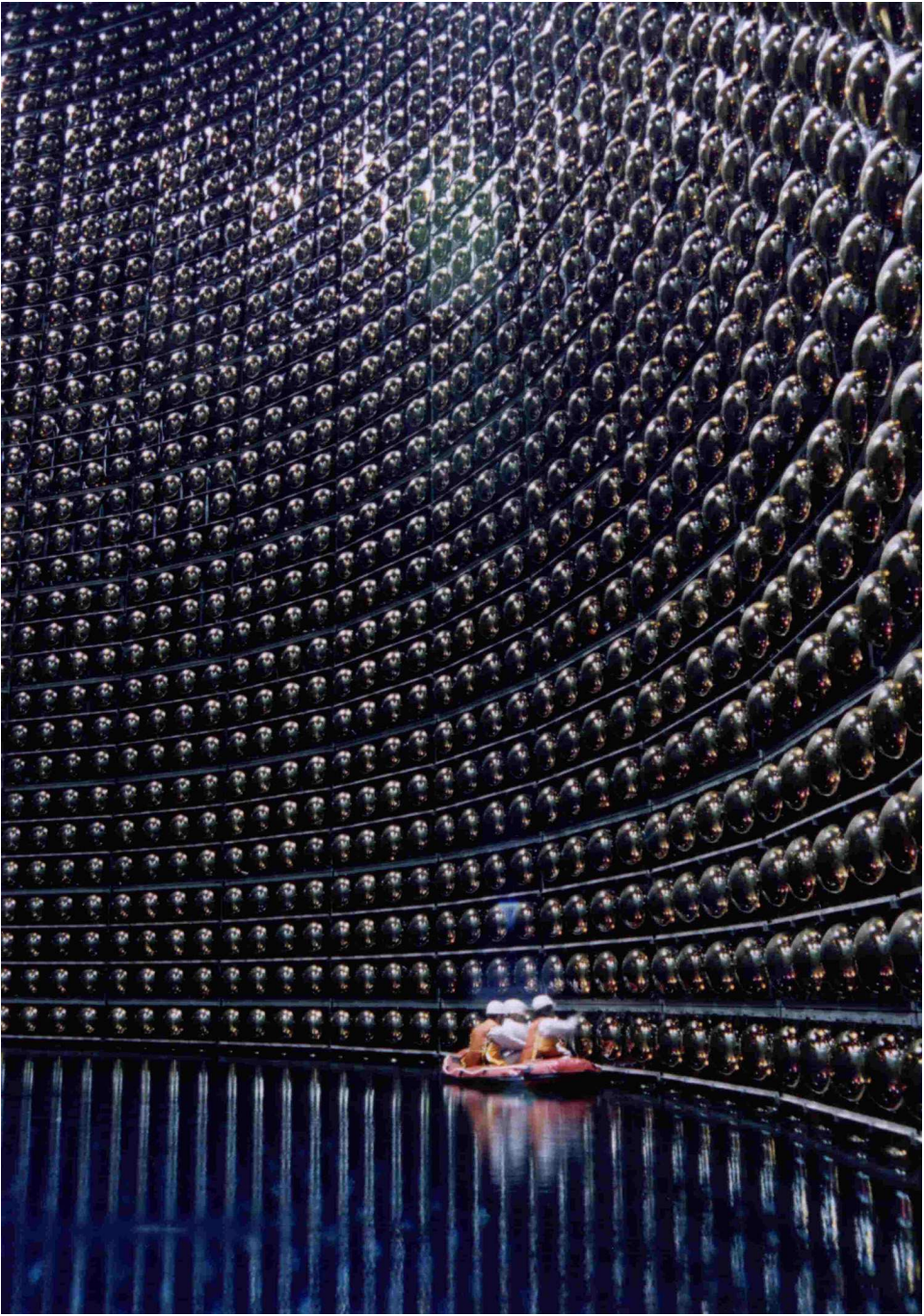}
 \caption{Super-Kamiokande is a 50,000 ton water tank surrounded by 11,146 of 20 inch PMTs (Photo-Multiplier Tubes). About half of the tank is filled with pure water. People on the boat are cleaning PMTs.  Credit: Kamioka Observatory, ICRR, Univ. of Tokyo \label{Fig4-neutrino.eps}}
 \end{figure}

\begin{figure}[htbp]
\centering
 \includegraphics[bb=0 0 360 150, width=5in]{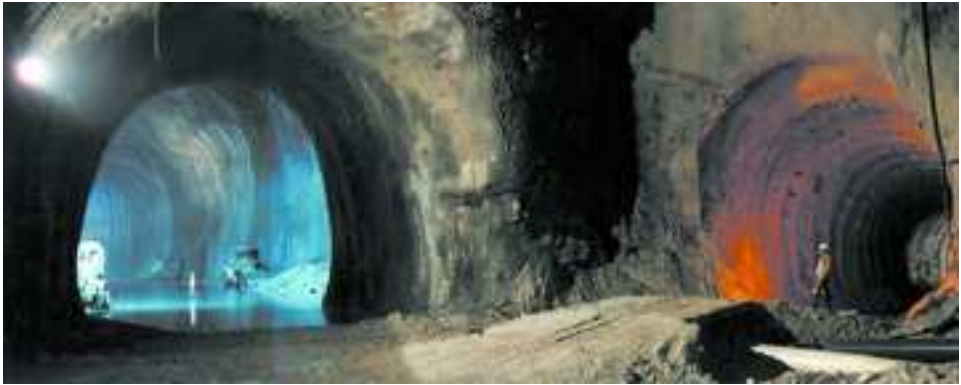}
 \caption{Excavations for National Laboratories of Gran Sasso. Deep underground experiments at sites like this are shielded by the earth above from most background effects, and are complementary to accelerator experiments at laboratories like the LHC. Credit: INFN \label{Fig5-IN0035H.eps}}
 \end{figure}

\begin{figure}[htbp]
\centering
 \includegraphics[bb=0 0 360 220, width=5in]{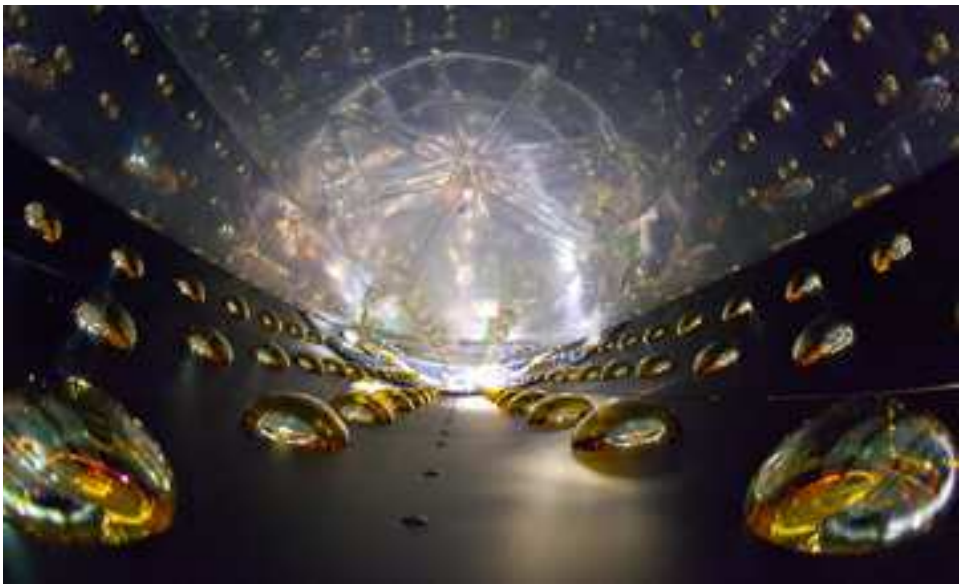}
 \caption{Photomultiplier tubes lining the walls of the Daya Bay neutrino detector in China, where the neutrinos from six nuclear reactors have been employed to clarify neutrino oscillations and masses. Credit: Roy Kaltschmidt, Lawrence Berkeley National Laboratory \label{Fig6-dayabay-hr.eps}}
 \end{figure}

\begin{figure}[htbp]
\centering
\includegraphics[bb=0 0 360 360, width=5in]{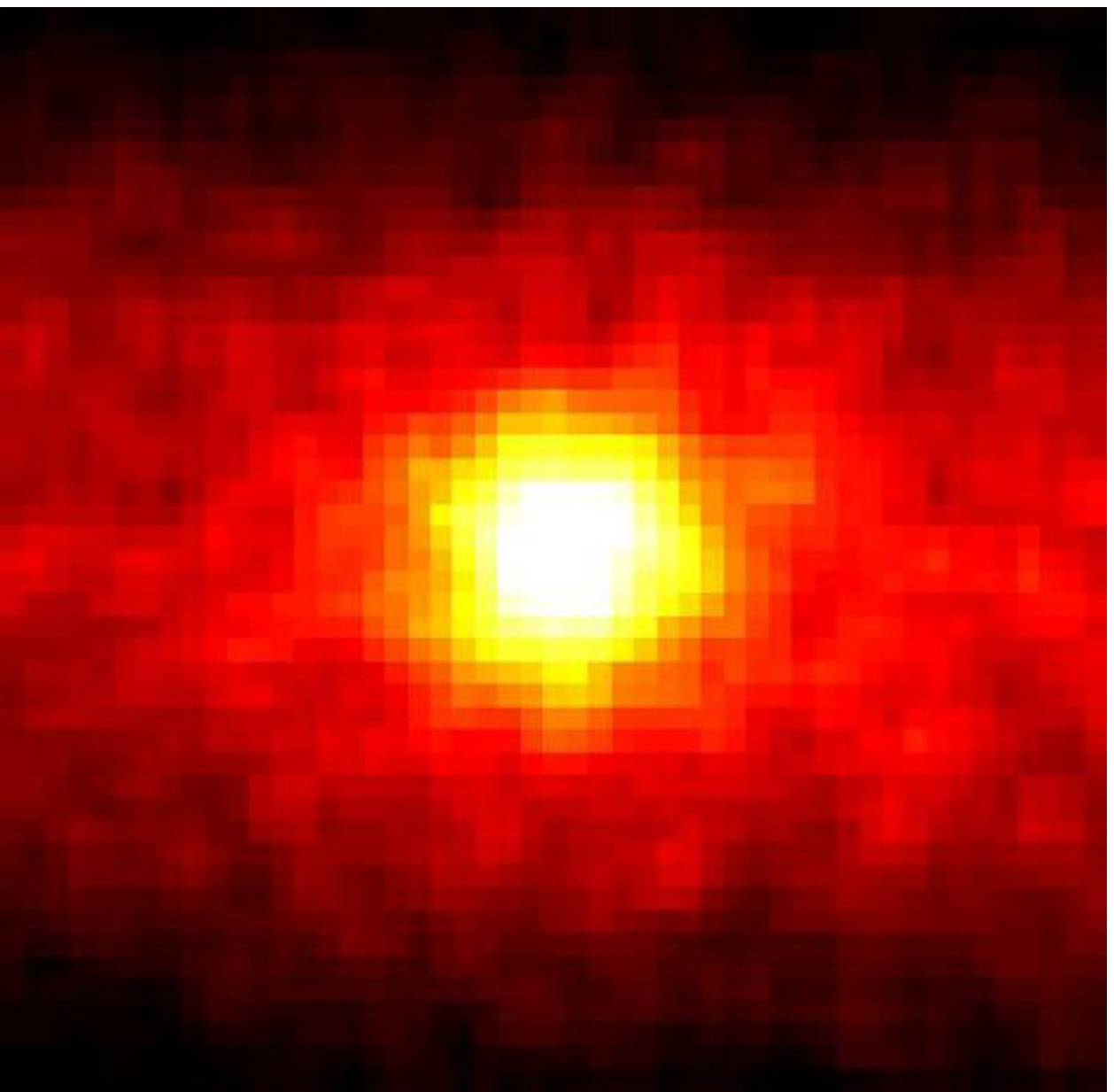}
\caption{The Sun as observed through its neutrino emissions by Super-Kamiokande. The first evidence of neutrino masses came from the heroic experiment of Ray Davis, which operated continuously from 1970 until 1994, and the parallel theoretical calculations of John Bahcall. When protons fuse to form $^4$He nuclei through a series of  reactions, electron neutrinos $\nu_e$ are emitted. Bahcall's calculations showed that the emitted number of $\nu_e$ should be three times the number of $\nu_e$ measured by Davis. The amazing fact was that this was not due to a failure of the experiment or an incorrect model of reactions deep inside the Sun. It instead resulted from conversion of the $\nu_e$ into all three of the neutrinos in Fig.~\ref{Fig1-FN0266H}, with this mixing being a result of their having masses. Credit: R. Svoboda, UC Davis, Super-Kamiokande Collaboration \label{Fig7-Sun-neutrinos.eps}}
\end{figure}

Here we consider the origin of particle masses within the SM, leaving neutrinos for the next section. In the SM, neutrinos have zero mass, but a number of heroic experiments, some of which are featured in Figs.~\ref{Fig3-FN0197H.eps}-\ref{Fig7-Sun-neutrinos.eps}, have now proved that neutrinos do have small masses. 

Various analogies have been offered for how the Higgs field gives mass to a matter particle. In each of these, the analogous effect is primarily relevant to velocity, whereas inertial mass is instead resistance to acceleration. But the spirit is correct. The analogy of John Ellis is to motion over a snow field, which represents the Higgs field. People wearing skis, snowshoes, or boots are respectively analogous to neutrinos, electrons, or top quarks, with little or no coupling to the Higgs field, moderate coupling, or strong coupling. Another analogy is motion through a field of party-goers. A nondescript stranger, a charming person like the reader of this article, and a celebrity like Paul McCartney will receive different degrees of attention from the crowd, and their attempts to move across the room will be impeded in proportion to their popularity.

A readable description of the research, and references to recent popular books on this general subject, can be found in the Popular Science Background published by the Royal Swedish Academy of Sciences~\cite{popular}. For example, the books by Frank Close, \textit{The Infinity Puzzle}, and Sean Carroll, \textit{The Particle at the End of the Universe}, are particularly timely and readable.

In the following, the matter and force particles of Fig.~\ref{Fig1-FN0266H} will be called by their conventional names, respectively fermions and vector (spin 1) bosons, with a Higgs being a scalar (spin 0) boson. It is assumed throughout this article that a reader who wants to understand the technical terms that are italicized, like \textit{fermion} above and \textit{propagator} below, can find their explanations by searching on the internet. We will also refer to equations by their numbers, such as (\ref{eq1.1}). 

\textbf{At this point, the reader who wishes to avoid the more technical aspects can skip the remainder of this section, plus the section immediately following, and go straight to Section \ref{sec:sec4}, on supersymmetry and dark matter. The discussion between here and Section \ref{sec:sec4} assumes some familiarity with complex numbers, vectors, and matrix multiplication.}

References to the detailed technical literature can be found in the proceedings of the recent Nobel Symposium on LHC results~\cite{Symposium}, in the Scientific Background on the Nobel Prize in Physics 2013~\cite{scientific} written by the Class for Physics of the Swedish Academy, and in reviews published by the Particle Data Group~\cite{pdg} . Regarding the connection to further discoveries of the future, there are many theories on every topic, and it would be prohibitively complicated to discuss them all. So the discussion here will be limited to specific mainstream scenarios, which provide a general guide to current thinking on these issues. It is also not feasible to properly credit those who made the discoveries and originated the ideas discussed here, but detailed citations are given in reviews like those of Ref.~\cite{pdg}.

\begin{figure}[htbp]
\centering
 \includegraphics[bb=0 0 120 540, width=1.2in]{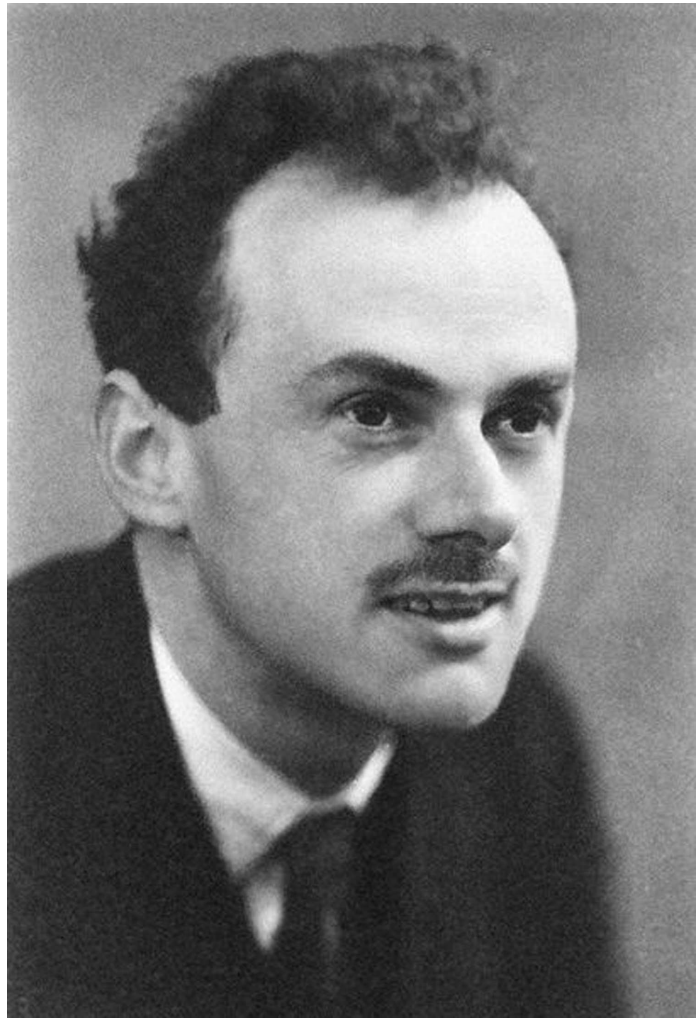}
 \caption{Paul Dirac, who provided the relativistic equation for electrons, which involves the mass term (\ref{eq2.1a}). It turns out that (\ref{eq2.1a}) ultimately leads to $m_{e}$ being the mass of an electron in the usual sense. In the SM, all the matter particles of  Fig.~\ref{Fig1-FN0266H} have these Dirac masses, except for the neutrinos. Credit: Nobel Foundation \label{Fig7a-Dirac.eps}}
 \end{figure}

We will now skip past a great deal of mathematics, and briefly outline 85 years of historical developments: After the introduction of quantum mechanics and Einstein's special theory of relativity, these two aspects of modern physics were combined in 1928 by Paul Dirac, shown in Fig.~\ref{Fig7a-Dirac.eps}. (So far no one has convincingly combined quantum mechanics with the general theory of relativity, which is essentially Einstein's theory of gravity, although for 30 years there has been a vigorous program with this goal called string theory.) In the general version of Dirac's theory, the complete field associated with a given fermion in Fig.~\ref{Fig1-FN0266H} consists of a left-handed field $\psi _{L}$ and a right-handed field $\psi _{R}$. The fermion mass $m_{f}$ then couples $\psi _{L}$ and $\psi _{R}$ through a term
\begin{eqnarray}
\psi _{L}^{\dag }\,m_{f}\,\psi _{R} \; .
\label{eq2.1}
\end{eqnarray}
These 2-component \textit{Weyl fields} transform oppositely under a \textit{Lorentz transformation} in relativity, and specifically under the \textit{Lorentz boost} of a particle to higher velocity, and this property is required to keep the whole expression constant under such a transformation. In the special case of the electron,  (\ref{eq2.1}) is
\begin{eqnarray}
e_{L}^{\dag }\,m_{e}\,e_{R} \; .
\label{eq2.1a}
\end{eqnarray}

\begin{figure}[htbp]
\centering
\fbox{\includegraphics[bb=0 0 260 190, width=5in]{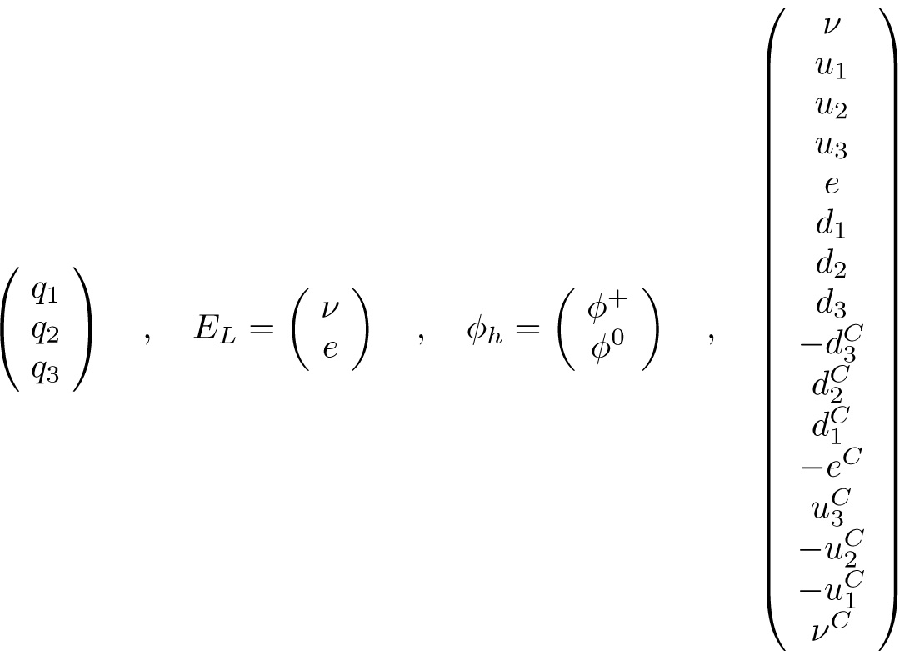}}
 \caption{Fields discussed in the text. In quantum physics, all of Nature consists of fields. Particles are excitations, or quanta, of these fields. A particular quark field comes in three \textit{colors} which are represented by $q_1$, $q_2$, and $q_3$ here, but which are conventionally called blue, green, and red. (These terms merely serve as labels for the quark states, of course, and are unrelated to the colors of light in ordinary experience.) The Higgs field $\phi_h$ of the SM actually consists of two fields, as shown. 
\newline \indent
In one candidate for a grand unified theory (GUT) of forces and particles, all of the matter fields of the SM are placed in three 16-component arrays like the one shown, representing the three generations of matter particles in Fig.~\ref{Fig1-FN0266H}. The one shown here is for the first generation -- the electron neutrino $\nu$, up quark $u$, electron $e$, and down quark $d$. When the bottom 8 fields are transformed from left-handed antiparticle fields (indicated by the superscript ``C'' for \textit{charge conjugate}) to right-handed fields for particles, each particle has both a left-handed and a right-handed field. For example, the electron has a left-handed field $e_L$ (called $e$ here) and a right-handed field $e_R$ (derived from $e^C$). 
\newline \indent
For the electron, the up and down quarks, and their relatives in the second and third generations, there is a single \textit{Dirac mass}, which is derived from the Higgs field. But the array shown contains one field beyond those of the SM --  an extra neutrino field. And the neutrino fields are permitted to have two kinds of masses, because they have no electric charge. They may have a Dirac mass, or a \textit{Majorana mass}, or both. 
\newline \indent
In one picture, the exotic right-handed neutrino field gets an enormous Majorana mass (from a Higgs-like field at extremely high energy), and it passes some of this along to the ordinary left-handed  neutrino. But the Majorana mass of the ordinary neutrino is very small, because of a \textit{seesaw mechanism}. So if the left-handed and right-handed fields of electrons and quarks are analogous to your left and right hands, the neutrino fields in this picture are more analogous to a lobster with an enormous right claw and tiny left one. Credit: This specific ordering of fields is taken from C. Kounnas, A. Masiero, D. V. Nanopoulos, and K. A. Olive, \textit{Grand Unification with and without Supersymmetry and
Cosmological Implications} (World Scientific, 1984). \label{Fig7b-fields.eps}}
 \end{figure}

But Lorentz invariance is not the only symmetry required. An expression like (\ref{eq2.1}) should also be invariant under a \textit{gauge transformation}, in which both the matter fields and the force fields, described by \textit{vector potentials} $A_{\mu}$, are each transformed in a specific way. It is outside the scope of this article to define precisely what this means, but the basic idea is that the matter fields (represented by column arrays in Fig.~\ref{Fig7b-fields.eps}) are rotated, in the same way that a vector (represented by its components) might be rotated. Then the force fields are also transformed, in such a way that the joint effect of transforming matter fields and force fields exactly cancels in a physically valid fundamental theory.

\begin{figure}[htbp]
\centering
\includegraphics[bb=0 0 100 600, width=1.0in]{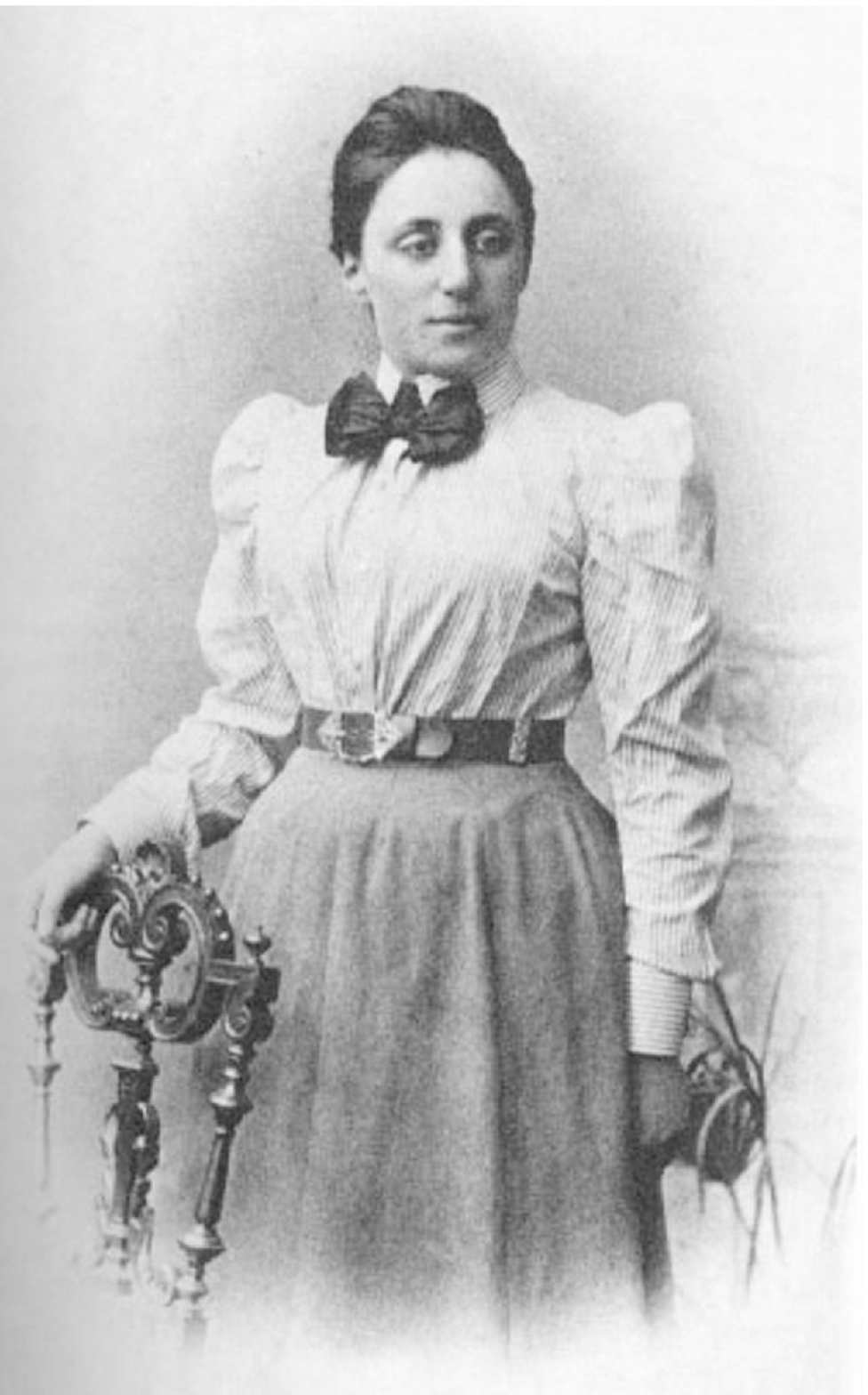}
\caption{Emmy Noether, who proved that symmetries imply classical conservation laws. The first version of Noether's theorem was published in 1918, and it demonstrated that energy was conserved in Einstein's theory of gravity, despite the initial doubts of mathematicians. Credit: unknown (Wikipedia) \label{Fig21-Emmy-Noether.eps}}
\end{figure}

A gauge transformation is purely mathematical, with no physical consequences. But if the theory is not invariant under gauge transformations, the conservation laws for related quantities will be broken. This follows from the connection between symmetries and conservation laws that was established by Emmy Noether, shown in Fig.~\ref{Fig21-Emmy-Noether.eps}. For example, the fact that the fundamental laws of Nature are the same at all times, the same at all positions in space, unaffected by rotation, and unaffected by electromagnetic gauge transformations implies conservation of energy, momentum, angular momentum, and electric charge. So Benjamin Franklin, the American founding father and scientist who discovered conservation of charge, would undoubtedly have been fascinated by the idea of gauge invariance.

To understand what this means for matter fields, first consider the $U \left( 1 \right)$, $SU \left( 2 \right)$, and $SU \left( 3 \right)$ gauge descriptions within the SM. ($SU \left( n \right)$ is the \textit{special unitary group} of $n \times n$ matrices, and $U \left( n \right)$ is the \textit{unitary group}.)  In each case, a gauge transformation essentially corresponds to a rotation of matter fields. In the $SU \left( 3 \right)$ description of the strong nuclear force, known as quantum chromodynamics, the fields being rotated have the form at the left of Fig.~\ref{Fig7b-fields.eps}: three quark fields labeled 1, 2, 3 which are assigned the \textit{colors} blue, green, and red.  So an $SU \left( 3 \right)$ gauge transformation rotates the quark fields of various colors into one another. 

There are eight gluon fields that do this, with the full vector potential being $A^i_{\mu} t^i$, and a sum implied over repeated indices like $i=1, 2 ,..., 8$. The $t^i$ are \textit{generators}, which start out as operators but are matrices in a given representation of the symmetry group. In the present context the generators are $3 \times 3$, $2 \times 2$, or $1 \times 1$ matrices, with the last just being complex numbers. They act on 3, 2, or 1 component fields. 

Passing on to $SU \left( 2 \right)$, we have $2 \times 2$ matrices and 2 component fields like those in the middle of Fig.~\ref{Fig7b-fields.eps}, representing the left-handed fields for the first generation of leptons in Fig.~\ref{Fig1-FN0266H} (the neutrino and electron), plus the Higgs field. In this case there are three vector bosons, which after Higgs condensation are rearranged to form the W$^+$, W$^-$, and Z$^0$ -- the carriers of the weak nuclear force. 

A single complex field $e_R$, the right-handed field for the electron, is typical for the fundamental $U \left( 1 \right)$ theory. There is also a single vector boson, which after Higgs condensation is combined with one of the original $SU(2)$ fields, the combination then being rearranged to form the photon and $Z^0$ fields. The rotation of $e_R$ under a gauge transformation is simply a rotation of this single complex field in the complex plane.

All these fields occur in the SM. But let us now jump to a leading candidate for a grand unified theory, shown at the extreme right of Fig.~\ref{Fig7b-fields.eps}. This \textit{spinor representation} of $SO \left( 10 \right)$, the group of rotations in 10 dimensions, has 16 fields. (The vector representation has 3 components in 3 dimensions and 10 in 10  dimensions, but the spinor representations turn out to have 2 and 16 respectively. $SO \left( n \right)$ is the \textit{special orthogonal group} of $n \times n$ matrices.)

Of course, the $SO \left( 10 \right)$ gauge theory is still formulated in four-dimensional spacetime, even though the gauge group is mathematically the same as the group of rotations in 10 dimensions. A group this large is required to include all the forces of the SM. The 16 fields in Fig.~\ref{Fig7b-fields.eps} contain all the fermions in one generation of Fig.~\ref{Fig1-FN0266H}: the electron $e$, the neutrino $\nu$, and the up and down quarks $u$ and $d$ in the three colors (labeled 1,2, 3 for blue, green, red). 

All of the 16 fields shown are left-handed (and listed and labeled in one of the possible conventions), but charge conjugation -- the transformation into a field with opposite \textit{quantum numbers} -- turns $e^{C}$ into the right-handed electron field, and similarly for each of the other $\psi ^C$. 

In other words, the original 8 left-handed fields of Fig.~\ref{Fig7b-fields.eps} for antiparticles turn into the 8 right-handed fields needed for the particles of Fig.~\ref{Fig1-FN0266H}, with each quark counted three times for colors. Now, however, there is one more field than in the SM, where neutrinos are strictly left-handed: In the grand unified $SO(10)$ theory we have acquired a right-handed neutrino field.  

Any fundamental gauge theory of the kinds described above does not permit fundamental fermion masses, because (\ref{eq2.1}) would then violate \textit{gauge invariance}: The fields $\psi _{L}$ and $\psi _{R}$ have different quantum numbers and therefore transform differently. 

Furthermore, fundamental masses for the vector bosons also violate gauge invariance: A mass term turns out to have the form
\begin{eqnarray}
M^{2}A^{\mu }A_{\mu }
\label{eq2.3}
\end{eqnarray}
and this would change when $A_{\mu }$ is transformed. (In relativity, the electromagnetic field is described by a \textit{four-vector} $A_{\mu }$, $\mu=0,1,2,3$,  which then gives rise to the electric field $\mathbf{E}$ and magnetic field $\mathbf{B}$. The same is true for the strong and weak nuclear forces. The expression $A^{\mu }A_{\mu }$, with an implicit sum over $\mu$, is analogous to the square of the length of a vector in three dimensions.)

The brilliant inventors of the SM therefore postulated that all particles are massless in the fundamental theory~\cite{Weinberg,Salam}, but that there are terms in which the fields interact with one another, including
\begin{eqnarray}
\lambda _{e}\,E_{L}^{\dag }\,\phi _{h}\,e_{R}\qquad ,\qquad E_{L}^{\dag}=\left( 
\begin{array}{cc}
\nu _{L}^{\dag } & e_{L}^{\dag }
\end{array}
\right) 
\label{eq2.4} \\
g^{2}\phi _{h} ^{\dag }A^{\prime \mu }\,A_{\mu }^{\prime }\,\phi _{h}
\label{eq2.5} \; .
\end{eqnarray}
We have allowed for the fact that the final physical fields $A_{\mu }$ can differ from the initial fundamental fields $A_{\mu }^{\prime }$. For example, the initial $U(1)$ field is different from the final $U(1)$ photon field, and the fundamental $U(1)$ coupling constant is different from the $U(1)$ coupling constant of electromagnetism (which corresponds to the electric charge $e$).

When the Higgs field condenses, it acquires a \textit{vacuum expectation value} 
\begin{eqnarray}
\left\langle \phi _{h}\right\rangle =\,\left( 
\begin{array}{c}
0 \\ 
\frac{1}{\sqrt{2}}v
\end{array}
\right) 
\label{eq2.6} \; .
\end{eqnarray}

Then (\ref{eq2.1a}) immediately follows from (\ref{eq2.4}), and (\ref{eq2.3}) follows from (\ref{eq2.5}) after detailed mathematics that is beyond the scope of this treatment. Both terms exhibit an \textit{effective} violation of gauge invariance, and, as a result, an \textit{effective} violation of the conservation laws that follow from gauge invariance. In the present case the fundamental $U(1)$ conserved quantity is called \textit{weak hypercharge} $Y$, and the $SU(2)$ conserved quantity $T_3$ is called \textit{weak isospin}. Only the electric charge $Q=T_3 + Y$ is conserved among particles after the Higgs field condenses. 

However, it is a fundamental principle that the initial symmetries and classical conservation laws are not broken when the vacuum in high-energy physics or ground state in condensed-matter physics are included. In particular, the Higgs condensate absorbs whatever quantities are lost by the particles. 

Notice that the original Higgs field of Fig.~\ref{Fig7b-fields.eps} has two components, each of which is complex, so there are four independent real fields. But the condensate consists of only a single real field in (\ref{eq2.6}). The other three real fields have been \textit{eaten by the vector bosons} $W^+$, $W^-$, and $Z^0$, each of which thereby acquires mass.  

\section{\label{sec:sec3} Neutrino masses, and grand unification of matter and force fields}

Let us now turn to neutrinos, which have been discovered to have small but nonzero masses. For the reasons given below, the observation of these masses definitely demonstrates physics beyond the Standard Model, and apparently points toward the need for a grand unified theory like that on the right-hand side of Fig.~\ref{Fig7b-fields.eps}. 

It is possible that neutrinos have Dirac masses, like the other fermions of Fig.~\ref{Fig1-FN0266H}, with a left-handed neutrino of the SM coupled to a new right-handed neutrino like that resulting from $\nu ^{C}$ in Fig.~\ref{Fig7b-fields.eps}, in a term with the basic form of (\ref{eq2.1}):
\begin{eqnarray}
\nu_R^{\dag} m_D \nu_L \; .
\label{eq1.2}
\end{eqnarray}
(We recall that Dirac masses in the SM result from the Higgs field.)

\begin{figure}[htbp]
\centering
\includegraphics[bb=0 0 100 560, width=1.0in]{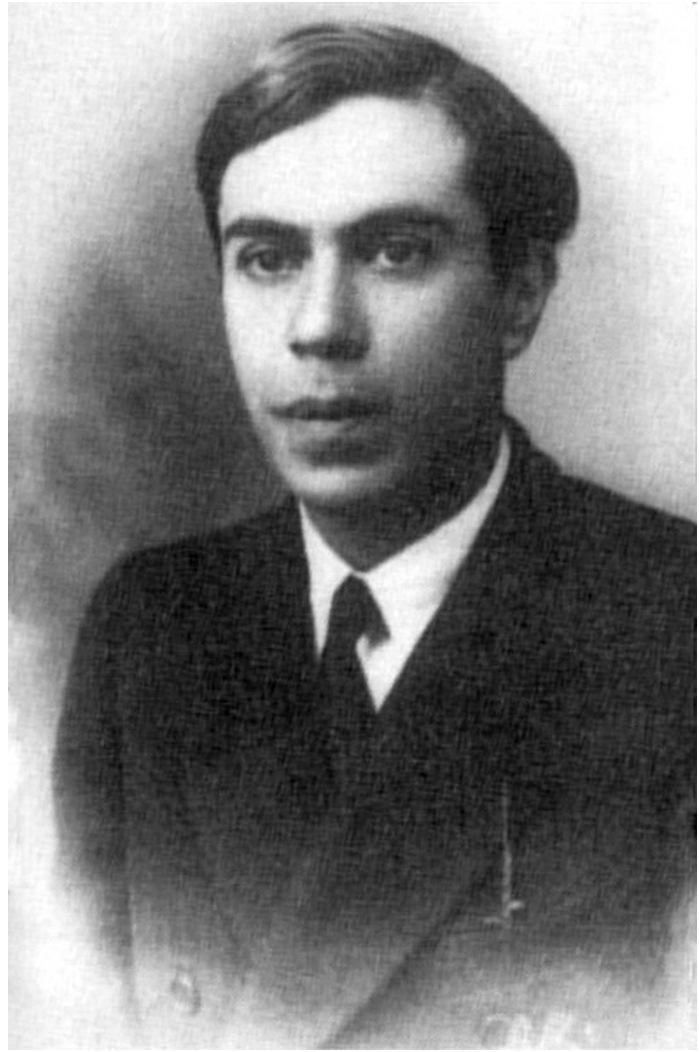}
\caption{Ettore Majorana, who pointed out the possibility of a second kind of matter particle mass, potentially relevant to both neutrinos and dark matter particles. Credit: unknown (Wikipedia) \label{Fig21a-EttoreMajorana.eps}}
\end{figure}

However, there is a second kind of particle mass, in which a particle's field is coupled not to another field, but essentially just to itself. This is called a \textit{Majorana mass}, named after the Italian physicist Ettore Majorana, shown in Fig.~\ref{Fig21a-EttoreMajorana.eps}.
In this case the field of a left-handed neutrino is again coupled to a right-handed field, but this other field is basically the charge conjugate of the original neutrino field, where charge conjugation essentially changes a particle into its antiparticle:
\begin{eqnarray}
\nu_L^C \, ^{\dag}  m_M \nu_L \; .
\label{eq1.3}
\end{eqnarray}

Again, it turns out that this form ultimately leads to a particle mass in the usual sense, even if a neutral particle has only one field rather than two. This is motivated by the fact that each neutrino in the SM does have only a left-handed field, unlike all the other matter particles, which have both left- and right-handed fields. It is also related to the fact that the weak nuclear force affects only the left-handed fields of all matter particles, including electrons and quarks.

The charge conjugate of $\nu_L$ has all the quantum numbers reversed in sign. Since the neutrino is assigned a \textit{lepton number} of +1, its charge conjugate has the opposite value -1, and a Majorana mass $m_{M}$ breaks lepton number conservation.

In beta decay for a neutron which is either free or inside an unstable atomic nucleus
\begin{eqnarray}
n \rightarrow p + e^- + \bar{\nu}
\label{eq1.4}
\end{eqnarray}
or the reaction associated with fusion in our Sun and other stars
\begin{eqnarray}
p \rightarrow n + e^+ + \nu
\label{eq1.5}
\end{eqnarray}
the neutrino $\nu$ and antineutrino $\bar{\nu}$ are normally regarded as distinct, with lepton number and baryon number conserved. (The neutron $n$, proton $p$, electron $e^-$, positron $e^+$, neutrino $\nu$, and antineutrino $\bar{\nu}$ respectively have baryon and lepton numbers of 1, 1, 0, 0, 0, 0 and 0, 0, 1, -1, 1, and -1.) 

\begin{figure}[htbp]
\centering
\includegraphics[bb=0 0 360 270, width=5in]{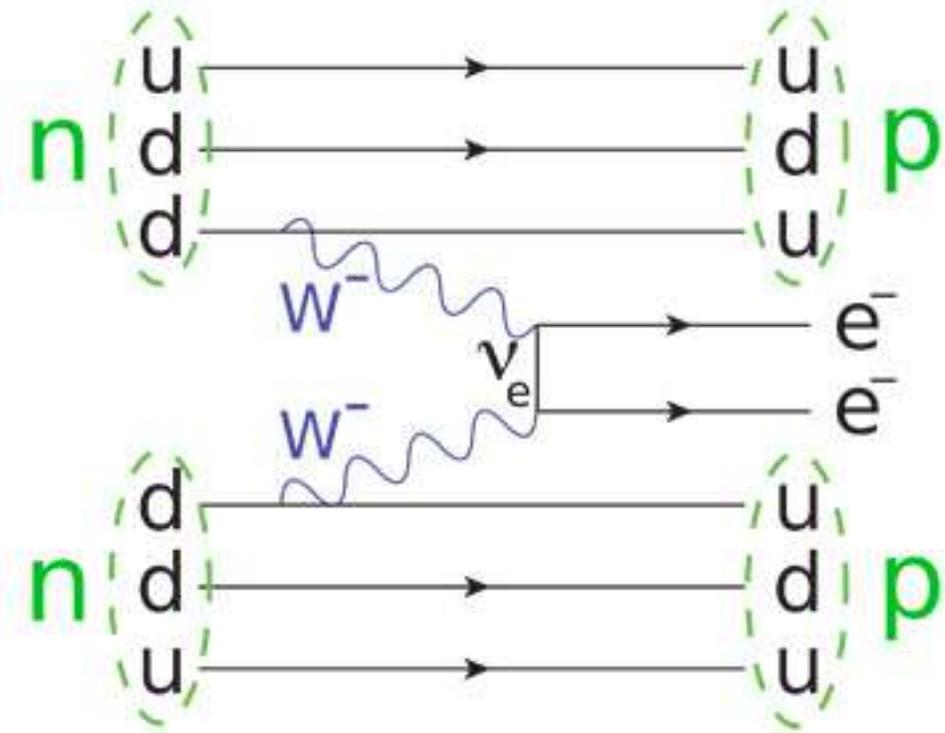}
\caption{Feynman diagram for neutrinoless double beta decay. Credit: JabberWok2 (Wikipedia) \label{Fig8-double-beta.eps}}
\end{figure}

On the other hand, if the neutrino mass is indeed Majorana, then the fact that such a mass breaks lepton number conservation will give rise to the very rare process of neutrinoless double beta decay, which is depicted in Fig.~\ref{Fig8-double-beta.eps}. 

This figure is an example of a \textit{Feynman diagram}, in which straight lines represent \textit{propagators} for fermions (matter particles), and wavy lines propagators for bosons (force particles). In the present case the fermions are quarks, electrons, and a neutrino, with the neutron $n$ composed of an up quark $u$ with charge $2/3$ and two down quarks $d$, each with charge $-1/3$.  (The proton is composed of two ups and one down.) A beta decay can still be represented by (\ref{eq1.4}), but also by 
\begin{eqnarray}
d \rightarrow u + e^- + \bar{\nu}
\label{eq1.6} \; .
\end{eqnarray}

However, a more fundamental description is the one shown in the figure, where the weak interaction first causes the process
\begin{eqnarray}
d \rightarrow u + W^-
\label{eq1.7} 
\end{eqnarray}
twice, followed by a process that might be pictured in different ways, with one being
\begin{eqnarray}
W^- \rightarrow e^-  + \bar{\nu} \label{eq1.7a}  \\
W^- + \nu \rightarrow e^-  \label{eq1.7b}  \; .
\end{eqnarray}
Then we have to interpret the $\bar{\nu}$ emitted in the first reaction to be the same as the $\nu$ absorbed in the second. 
We might pin down exactly how this can happen, by putting an $X$ in the middle of the $\nu_e$ line to indicate that it is the Majorana mass of (\ref{eq1.3}) that intervenes to change the -1 of $\bar{\nu}$ to the +1 of $\nu$. 

This can happen if, in a still more fundamental description, we replace (\ref{eq1.3}) by
\begin{eqnarray}
\lambda_{\nu} \nu_L^C \, ^{\dag}  \phi_{\nu} \nu_L 
\label{eq1.8a}
\end{eqnarray}
where $\lambda_{\nu}$ is a constant. (As will be seen below, a plausible picture derives (\ref{eq1.8a}) from an initially indirect coupling to $\phi_{\nu}$.) Then, if $\phi_{\nu}$ condenses, (\ref{eq1.8a}) becomes
\begin{eqnarray}
\nu_L^C \, ^{\dag}  \lambda_{\nu} \langle  \phi_{\nu} \rangle \nu_L 
\label{eq1.8}
\end{eqnarray}
which provides the mass of (\ref{eq1.3}). 

Once again, we see the centrality of the Higgs mechanism: It is directly connected to the second strong experimental indication of new physics, neutrino masses.

The field $\phi_{\nu} $ itself has quantum numbers, including a lepton number of -2, so that $\phi_{\nu} \nu_L$ and $\nu_L^C$ have the same quantum numbers in (\ref{eq1.8a}) and Fig.~\ref{Fig8-double-beta.eps}, and in the most fundamental picture lepton number is conserved. There is again an \textit{effective} violation of lepton number because the $\langle  \phi_{\nu} \rangle$ condensate absorbs whatever lepton number is missing for the particles depicted in Fig.~\ref{Fig8-double-beta.eps}. And again one normally discusses these processes within the effective theory, in which only particles appear and not the vacuum.

Furthermore, within the effective theory it turns out that one can even interpret the neutrino to be its own antiparticle. We might then reinterpret the $\nu_e$ line of Fig.~\ref{Fig8-double-beta.eps} as emission of a neutrino by one W$^-$ and subsequent absorption by the other W$^-$. Or we might even imagine  the $\nu_e$ line as representing particle-antiparticle annihilation of two neutrinos which are emitted from the two W$^-$ bosons. It is important to realize, however, that the mathematical expression corresponding to Fig.~\ref{Fig8-double-beta.eps} is unambiguous and not affected by the language that we use to describe this process. (To some extent, the language describing Feynman diagrams is metaphorical.)

There is an eminently plausible mechanism for the origin of a small Majorana mass, in which $\nu_L$ is coupled to another independent field $\nu_R$ (resulting from e.g. Fig.~\ref{Fig7b-fields.eps}) through a Dirac mass, which comes from a Higgs field in the same way as for electrons etc. in the preceding section. Then $\nu_R$ directly acquires a Majorana mass by being coupled to a GUT field $\phi_{\nu}$ which condensed in the very early universe, long before the Higgs field did so.

It follows that (1) $\nu_L$ is mixed with $\nu_R$ and thereby acquires a Majorana mass, but (2) this mass is very small because of a \textit{seesaw mechanism}: The mass of $\nu_L$ is pushed down by the large mass $M_R$ of $\nu_R$, since it turns out that 
\begin{eqnarray}
m_L \sim m_D^2/M_R \; .
\label{eq1.9}
\end{eqnarray}
The observed neutrino masses are, of course, small compared to the $m_D \sim m_e$ or even $m_D \sim m_{\tau}$ (mass of $\tau$ particle in Fig.~\ref{Fig1-FN0266H}) that might have otherwise been expected.

The various symbols in Feynman diagrams like Fig.~\ref{Fig8-double-beta.eps} correspond to precise mathematical expressions, with precise rules for how these expressions are arranged and used to calculate probabilities or rates for a process. The external lines correspond to real particles going into or coming out of the process, and the internal lines correspond to \textit{virtual} particles. These internal particles do not have the usual relation between energy $E$ and 3-momentum $\vec{p}$; instead each of these quantities freely ranges over all allowed values. But in other respects virtual particles have the same properties as real ones, including quantum numbers and masses. 

\section{\label{sec:sec4} Supersymmetry and dark matter}

To recap the foregoing discussion, there are two types of masses for fermions, like the matter particles on the left of Fig.~\ref{Fig1-FN0266H}: An electron has a Dirac mass, which connects its left-handed field to its right-handed field. A neutrino can have a Majorana mass, which connects a left-handed neutrino only to itself. A Dirac particle has an antiparticle which is different from itself; for example, the antiparticle of an electron is a positron. But a Majorana particle can be its own antiparticle.  

This idea spills over into the realm of supersymmetic fermions, which are depicted in Fig.~\ref{Fig9-susyparticles_sm.eps}. As described in the caption, susy provides an excellent dark matter candidate, called the neutralino. This is a spin 1/2 fermion with zero  charge. 

No sparticles have yet been seen, and they would have been if they had the same masses as their partners in the SM. So Nature must somehow have provided a mechanism for \textit{supersymmetry breaking}, which causes the masses of the neutralino and other sparticles to lie well above those of most SM particles. Supersymmetry breaking is not well understood, but there are various proposals which yet again involve condensation of some bosonic field, which plays essentially the same role as the Higgs field.

This is another example of the centrality of the Higgs phenomenon: In various forms, it shows up in the context of neutrino masses, grand unification, and supersymmetry, as well as in the original context of the SM.

\begin{figure}[htbp]
\centering
 \includegraphics[bb=0 0 360 300, width=5in]{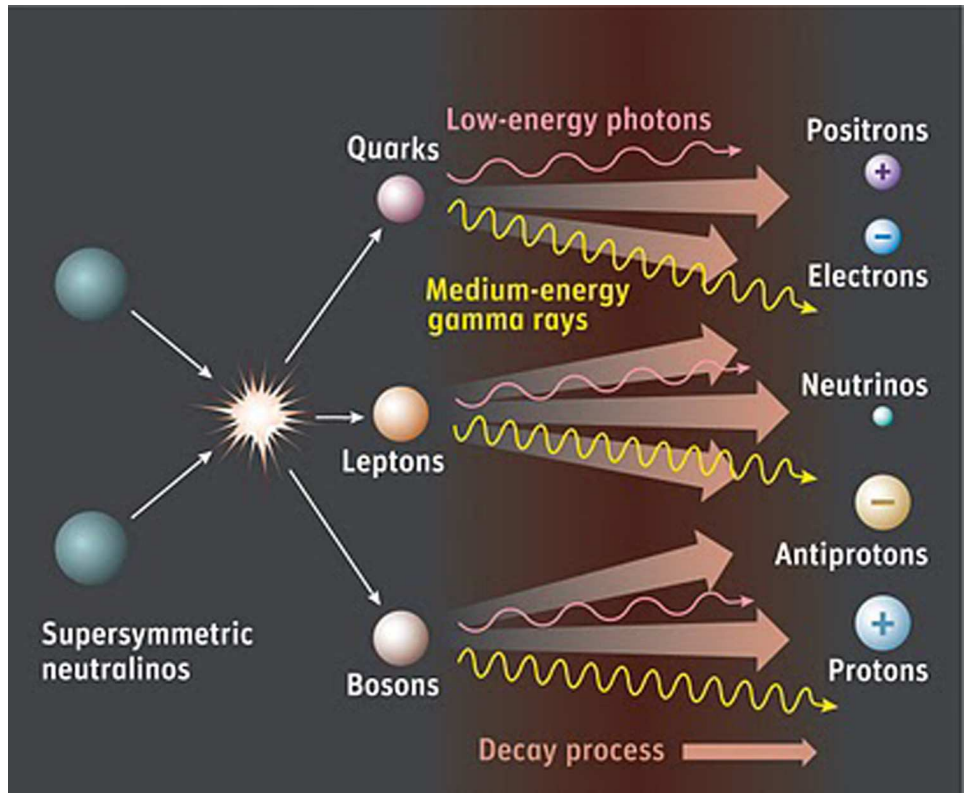}
\caption{Neutralino-neutralino self-annihilation, producing particles at a well-defined energy determined by the neutralino mass --  a signature that can be detected in satellite observatories like Fermi, PAMELA, and AMS. Credit: Alexander. B. Fry (www.theastronomist.com) \label{Fig10-neutralino.eps}}
\end{figure}

\begin{figure}[htbp]
\centering
 \includegraphics[bb=0 0 360 250, width=5in]{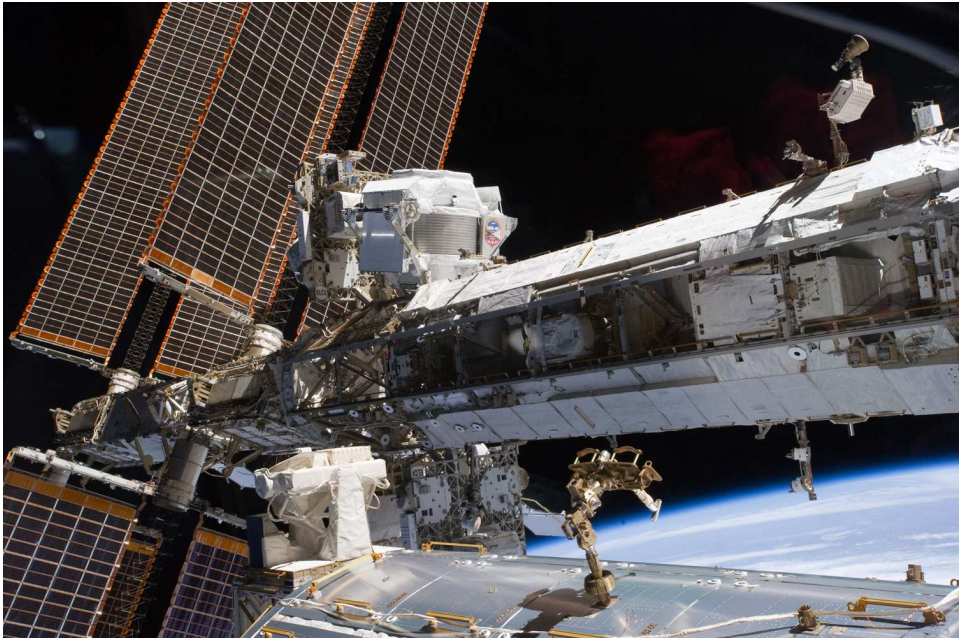}
 \caption{The International Space Station, photographed by a crew member while space shuttle Endeavour was docked with the station, on May 20, 2013. The newly-installed Alpha Magnetic Spectrometer-2 (AMS) is visible at center left. The Earth's horizon can be seen below. AMS, like the Fermi and PAMELA experiments, is looking for evidence of particles created through dark matter annihilation. Credit: NASA \label{Fig11-AMS-02.eps}}
\end{figure}

If the neutralino is a Majorana particle, it can be interpreted as its own antiparticle, and can therefore undergo the same kind of annihilation as in one way of picturing the process in Fig.~\ref{Fig8-double-beta.eps}. This time, however, the two annihilating particles are real rather than virtual, and real energy must emerge from the event, carried by real particles, as shown in Fig.~\ref{Fig10-neutralino.eps}. Evidence has been sought for this process in satellite experiments -- Fermi, PAMELA, and most recently AMS, which is depicted in Fig.~\ref{Fig11-AMS-02.eps} -- but the discovery of a definitive signal must await more data.

Astronomy provides strong evidence for dark matter in many different ways, going back to the observations of Fritz Zwicky in the 1930s (involving the motion of galaxies), continuing with those of Vera Rubin and her colleagues in the 1970s (involving the motion of stars around galaxies), and more recently involving detailed understanding of how galaxies and larger structures came to form. 

\begin{figure}[htbp]
\centering
 \includegraphics[bb=0 0 360 270, width=5in]{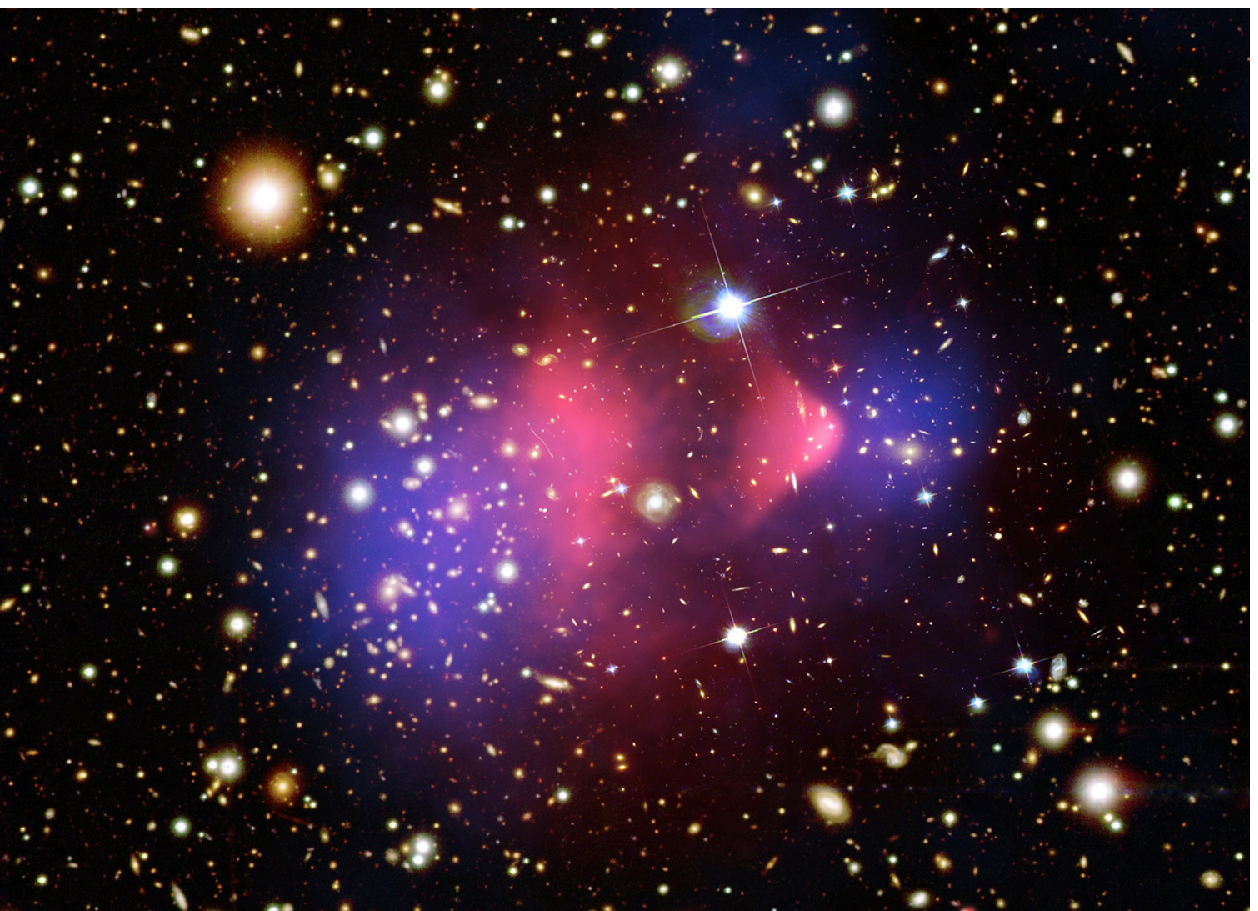}
 \caption{In this composite image of the bullet cluster (galaxy cluster 1E 0657-56), the red region shows the location of ordinary matter, as mapped out by hot gas. The blue region shows the gravitational mass, as mapped out by gravitational lensing. The interpretation is that, in a collision between two galactic clusters, the ordinary matter was slowed by drag forces. The dark matter does not experience these forces, and as a result it freely sails though, unimpeded by any force except gravity. Credit: X-ray: NASA/CXC/M.Markevitch et al. Optical: NASA/STScI; Magellan/U.Arizona/D.Clowe et al. Lensing Map: NASA/STScI; ESO WFI; Magellan/U.Arizona/D.Clowe et al. \label{Fig12-bulletcluster.eps}}
\end{figure}

Further evidence is provided by studies like those of Fig.~\ref{Fig12-bulletcluster.eps} using gravitational lensing, in which light from extremely distant sources is bent around objects that are closer. The amount that light is bent by the gravitational field in a region serves as a probe of the density of the matter in that region, both luminous and dark.

\begin{figure}[htbp]
\centering
\includegraphics[bb=0 0 360 360, width=5in]{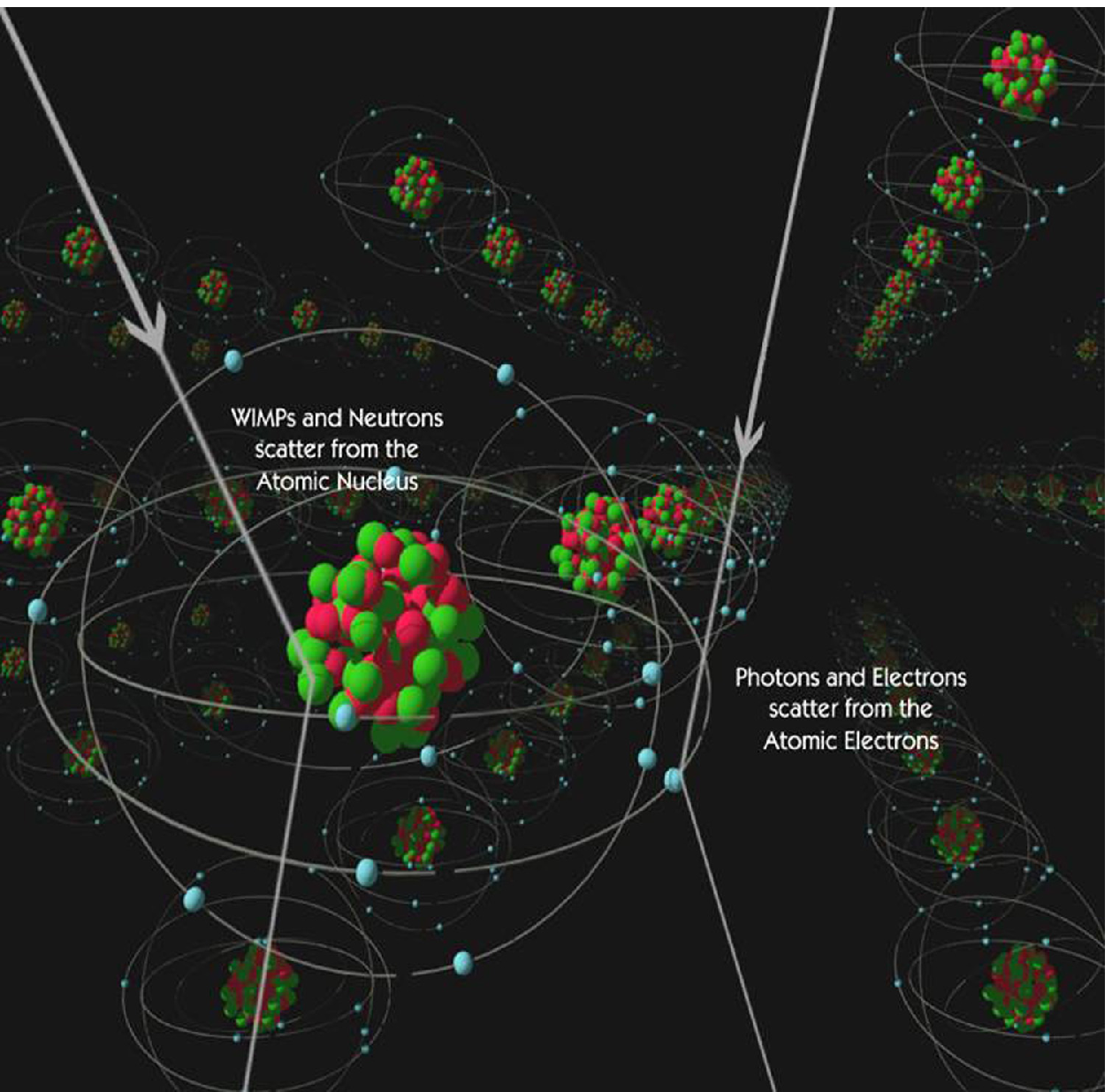}
\caption{Dark matter collisions can be distinguished from the background of other collisions because they involve the atomic nucleus rather than the electrons. Credit: SuperCDMS Collaboration \label{Fig13a-CDMS.eps}}
\end{figure}

Strenuous efforts are underway to directly detect dark matter through rare collisions with atomic nuclei in laboratories on the Earth. A basic  principle is shown in Fig.~\ref{Fig13a-CDMS.eps}: Collisions of dark matter particles can be distinguished from other events because these particles  do not experience the electromagnetic force, and as a result they collide with the atomic nucleus rather than electrons. 

\begin{figure}[htbp]
\centering
\includegraphics[bb=0 0 360 260, width=5in]{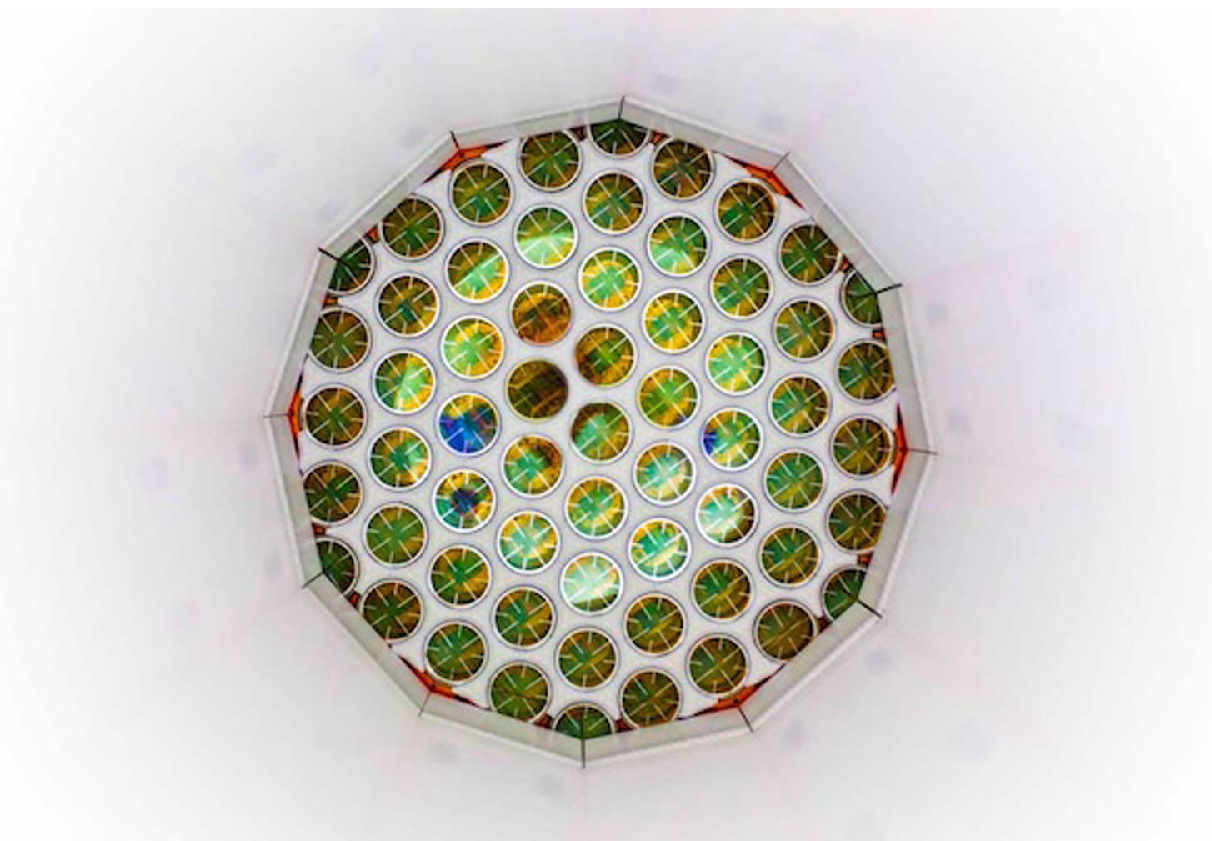}
\caption{Photomultiplier tubes in the LUX dark matter detector. They are capable of detecting as little as a single photon of light, resulting from collisions of dark matter particles with xenon nuclei. Credit: Matt Kapust, Sanford Underground Research Facility \label{Fig13x-LUX.eps}}
\end{figure}

Fig.~\ref{Fig13x-LUX.eps} shows photomultiplier tubes used in the Large Underground Xenon (LUX) experiment, which currently is the world's most sensitive. It employs 370 kilograms of liquid xenon, and is located about 1.5 kilometers underground, to shield it from background radiation at the Earth's surface. Its site, in the former Homestake Gold Mine in the Black Hills of South Dakota, was previously used in the solar neutrino experiment of Ray Davis, described in the caption to Fig.~\ref{Fig7-Sun-neutrinos.eps}.

\begin{figure}[htbp]
\centering
\includegraphics[bb=0 0 360 270, width=5in]{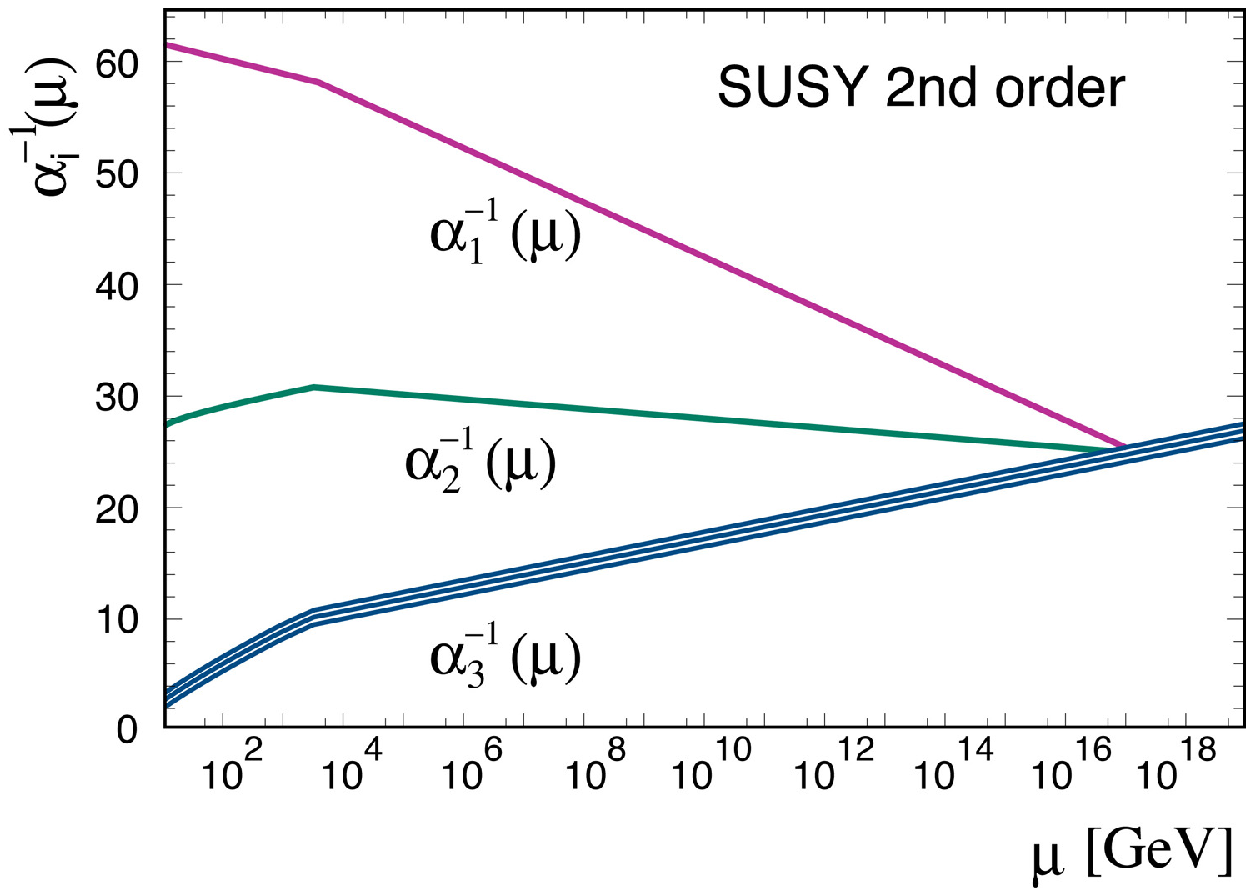}
\caption{The coupling constants of the nongravitational forces converge to a common value at the GUT scale if susy is included (whereas they fail to converge without susy). In this figure it is assumed that the susy particles have masses not far from 1 TeV. Credit: Jean-Pierre Revol, CERN \label{Fig13b-susy-GUT.eps}}
\end{figure}

There is already indirect experimental evidence for susy: The strength of each of the three fundamental forces in the Standard Model is characterized by a \textit{coupling constant}. In order for all three forces to have descended from a single grand-unified force, as the universe cooled following the Big Bang, it is necessary that the coupling constants converge to a common value at high energy. As Fig.~\ref{Fig13b-susy-GUT.eps} shows, this happens when susy is included. But the same graph without susy fails to show such a clean convergence.

\section{\label{sec:sec5} How the Higgs was discovered}

The LHC supports four experiments: ALICE (A Large Ion Collider Experiment) is designed to use Pb-Pb nuclear collisions to study the quark-gluon plasma, which existed in the early universe before these two kinds of partons froze out to form protons and neutrons. The LHCb (Large Hadron Collider beauty) experiment is designed to help explain why we live in a universe that appears to be composed almost entirely of matter, and almost no antimatter, by studying events involving the bottom quarks of Fig.~\ref{Fig1-FN0266H}, sometimes whimsically called beauty quarks. 

According to the SM, matter and antimatter should have totally annihilated each other in the early universe, leaving no matter to form stars, planets, and us. The criteria which permit the survival of matter were laid down by the eminent Russian scientist (and Soviet dissident) Andrei Sakharov. They require new physics beyond the SM.

\begin{figure}[htbp]
\centering
\includegraphics[bb=0 0 360 200, width=5in]{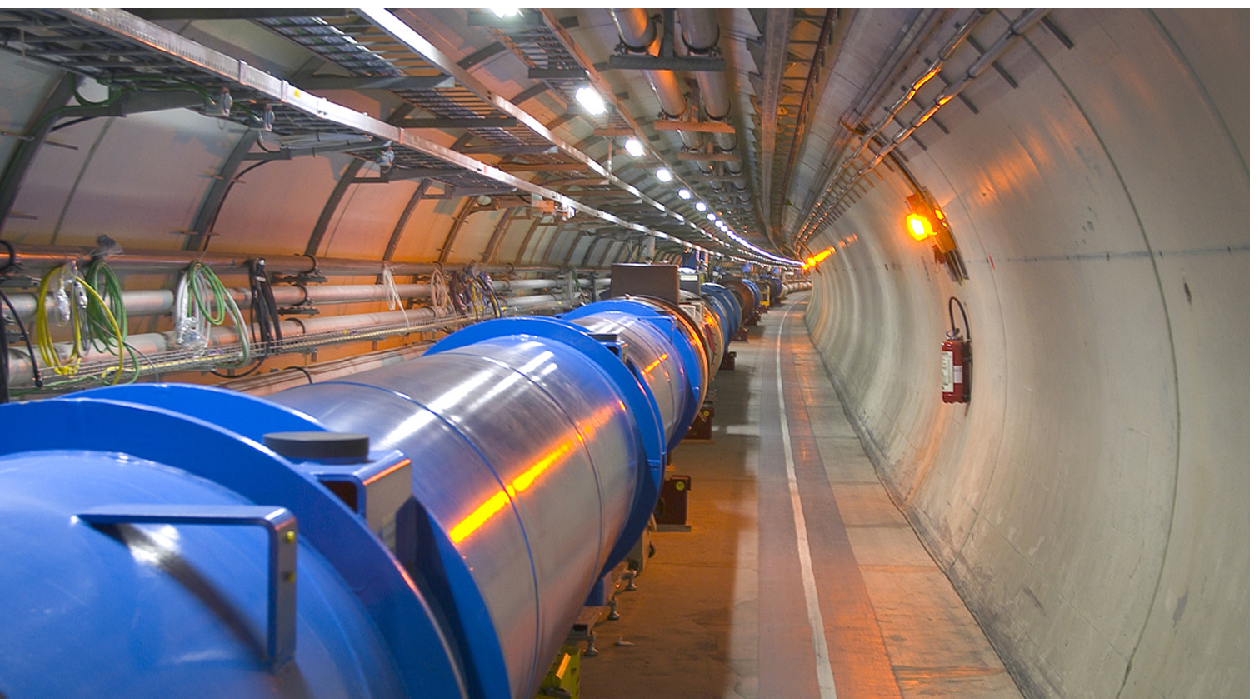}
\caption{Inside the LHC tunnel, deep underground. The two counter-rotating beams of protons (or heavy ions for ALICE) are made to bend by the large dipole magnets, and are kept extremely narrow by focusing magnets. Credit: CERN \label{Fig13-CE0085H.eps}}
\end{figure}

These are important issues, but the discovery of a Higgs boson was made by the CMS (Compact Muon Solenoid) and ATLAS  (A Toroidal LHC Apparatus) experiments, each of which is operated by a team of more than 3000 investigators. About 1/3 of the team members are graduate students, and about 1/4 are women. Lining  the 27 kilometers of tunnel are 1232 dipole bending magnets, shown in Fig.~\ref{Fig13-CE0085H.eps}, which employ coils of superconducting niobium-titanium cable. Each of the bending magnets is more than 14 meters long, has a mass of 35 metric tons, carries almost 12,000 amperes of current, creates a magnetic field of more than 8 tesla, has its superconducting cables cooled below 2 K, and is subjected to enormous mechanical stresses because of the powerful magnetic fields. The total energy stored in the magnets is about $11 \times 10^9$ joules. There are also quadrupole focusing and other specialized magnets. The beam diameter is about 1/3 the diameter of a human hair at the collision points. The beam vacuum, with a pressure of $10^{-13}$ atm, is comparable to that of outer space. 

\begin{figure}[htbp]
\centering
\includegraphics[bb=0 0 360 250, width=5in]{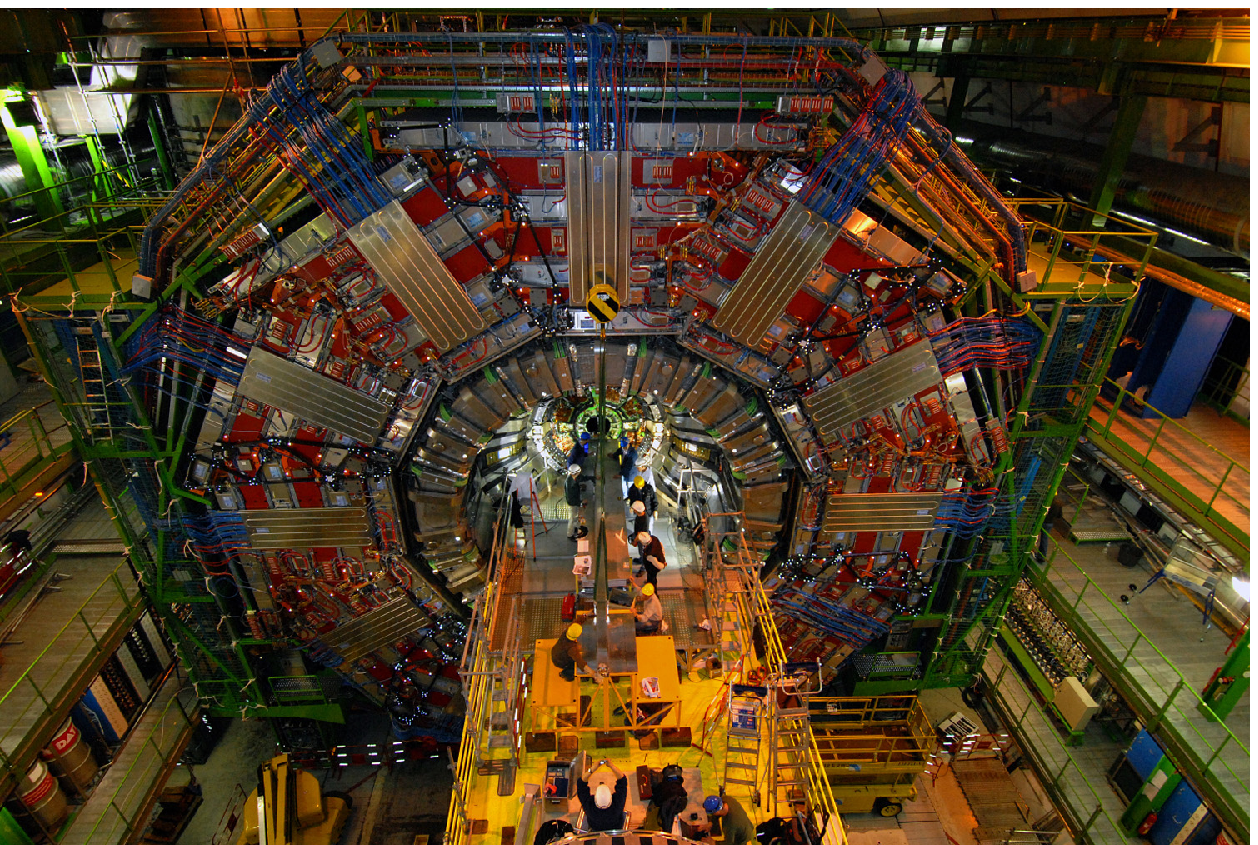}
\caption{CMS detector under construction. The scale is set by the human workers near the center. Credit: CERN \label{Fig14-DSC_0159.eps}}
\end{figure}

\begin{figure}[htbp]
\centering
\includegraphics[bb=0 0 360 270, width=5in]{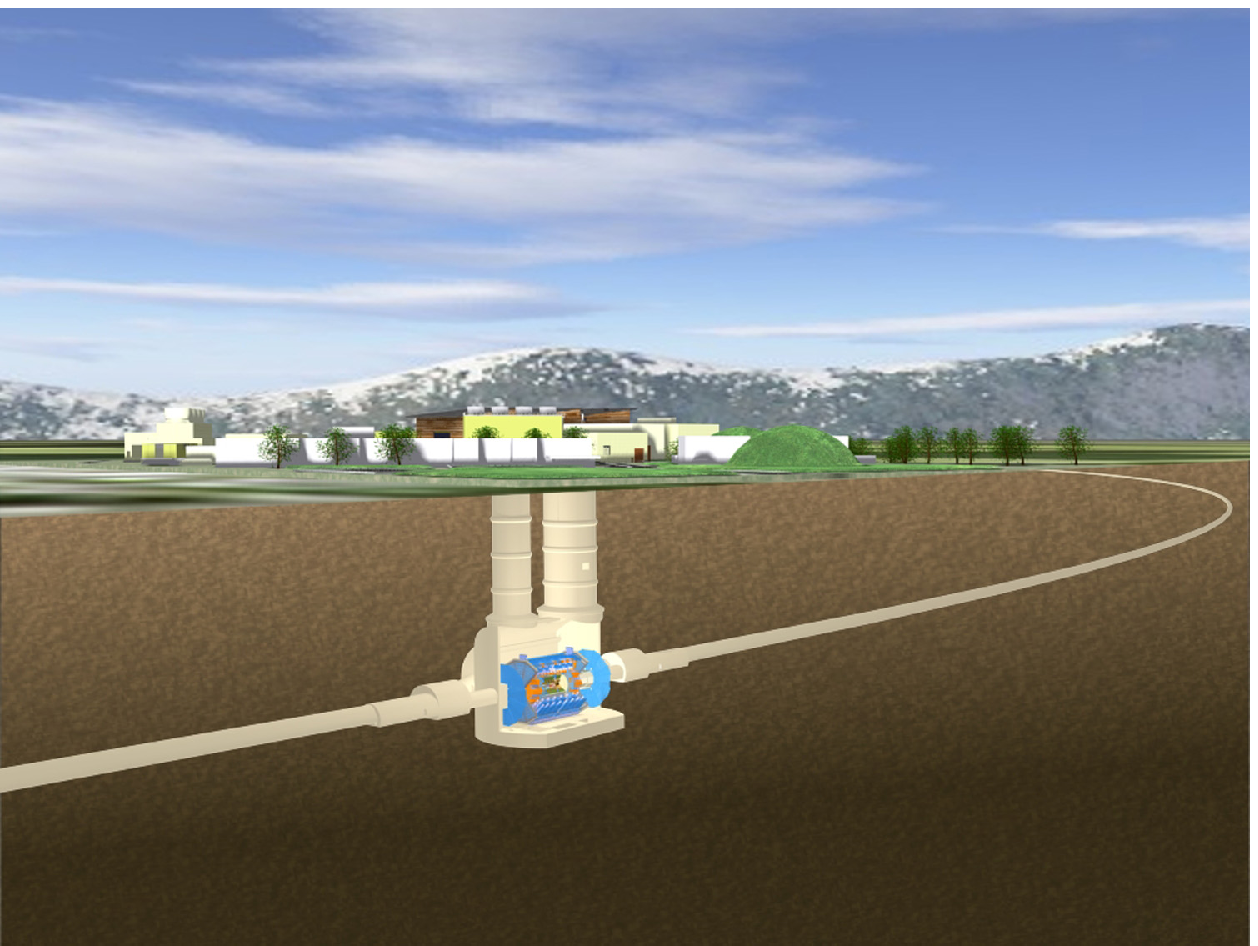}
\caption{Illustration of the ATLAS detector in its cavern 100 meters underground, with the LHC tunnel extending in either direction. Credit: CERN \label{Fig14a-atlasdet.eps}}
\end{figure}

The four experiments -- i.e. detectors -- are located at four different points around the ring of Fig.~\ref{Fig2-8701973.eps}. They must be large in order to capture the tracks of extremely energetic particles emitted from the collisions. But they must also be exquisitely precise for quantitative measurements. Figure~\ref{Fig14-DSC_0159.eps} shows the CMS detector when it was being constructed, and Figure~\ref{Fig14a-atlasdet.eps} illustrates the position of the ATLAS detector.

\begin{figure}[htbp]
\centering
\includegraphics[bb=0 0 360 230, width=5in]{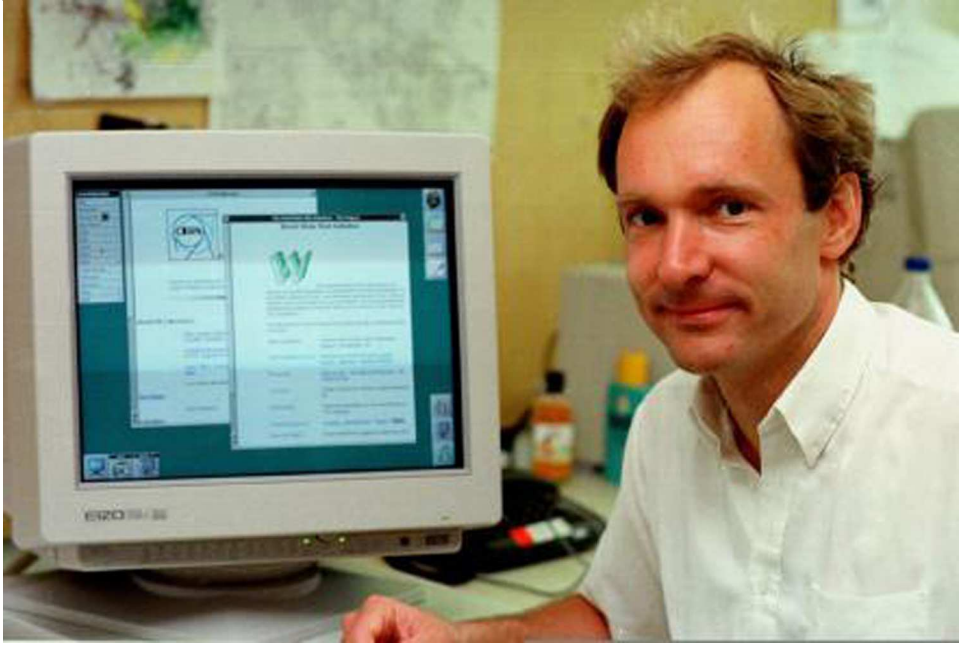}
\caption{Timothy Berners-Lee, who invented the World Wide Web at CERN, in order to permit large experimental collaborations to share their data. Credit: CERN \label{Fig15-Berners-Lee.eps}}
\end{figure}

Disentangling the results of the collisions is a task of mind-boggling difficulty, because there are about 20,000,000 collisions per second, and nearly all the emitted particles come from processes that are already thoroughly understood. The first step is to create software that stores only the results most likely to show new physics -- typically 300 events per second. Then these have to be analyzed. So there are three major requirements: an extremely powerful accelerator laboratory (the LHC), extremely sophisticated experiments (the detectors), and extremely powerful computing resources. Fortunately, those who work at CERN, Fermilab, and their sister laboratories have developed considerable skill at developing software and employing it to understand the results of collider experiments. As one example, the World Wide Web, which now makes the resources of the internet available in a vast number of applications (like Google and Facebook), was developed at CERN by Timothy Berners-Lee, shown in Fig.~\ref{Fig15-Berners-Lee.eps}.

There are several ways in which the discovery of the Higgs boson was more difficult than one might have expected, even given the enormous background of events mentioned above. First, it was found in the last of the most likely places, with a mass of 126 GeV when given in terms of its energy equivalent. (This means that $m_h c^2 = 126 \times 10^9$ eV, where 1 eV is the energy acquired by an electron or proton accelerated through 1 volt of electric potential. A typical atomic or chemical energy is a few eV, and a temperature of 1 K corresponds to about $10^{-4}$ eV. The Higgs boson mass is then about 130 times the mass of a proton.) The other favored energy regions, lower and higher, had already been ruled out for a Higgs consistent with the SM. Second, there is the ironic fact that one of the cleanest signatures of a Higgs is its decay into two high-energy photons, with this being a low-probability process. The higher-probability processes are less easily disentangled. 

The value of the Higgs mass $m_h$ obtained by CMS and ATLAS has a remarkable property (according to detailed calculations which rely on accurate measurement of the top quark mass): If $m_h$ were definitively lower, the Higgs condensate would be unstable, and if $m_h$ were definitively higher it would be stable. But $m_h$ appears to be very near the critical value where the Higgs condensate is bordering on instability~\cite{Degrassi}. The same is thus true of the universe as we know it. This rather unsettling fact may be a clue in the search for a more fundamental theory. 

\begin{figure}[htbp]
\centering
\includegraphics[bb=0 0 360 240, width=5in]{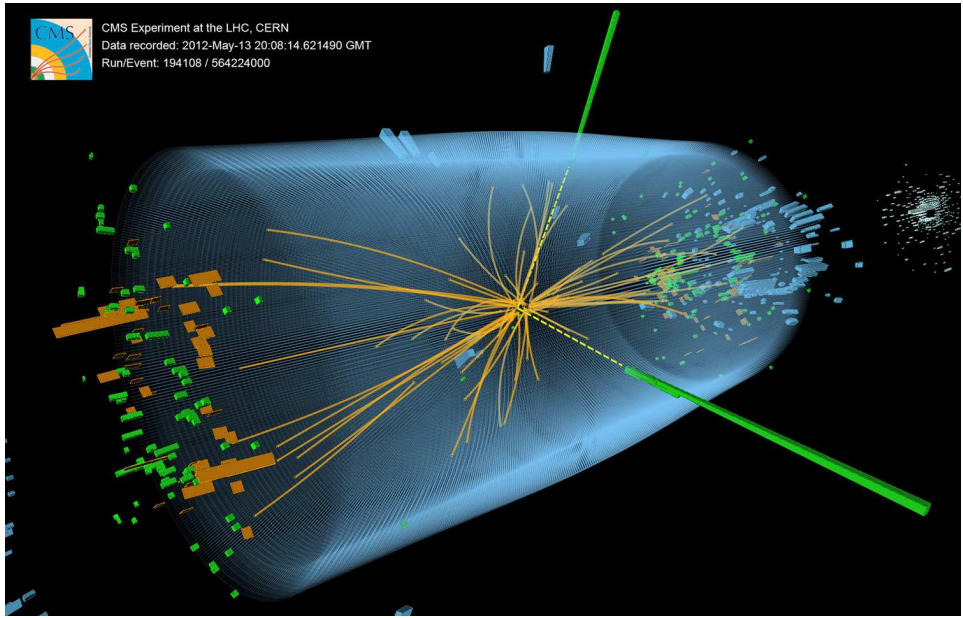}
\caption{A CMS candidate for a Higgs decaying into two photons, indicated by the two long (green) lines, plus other particles. Credit: McCauley, Thomas; Taylor, Lucas; CERN \label{Fig16-CMS-gammas.eps}}
\end{figure}

\begin{figure}[htbp]
\centering
\includegraphics[bb=0 0 360 280, width=5in]{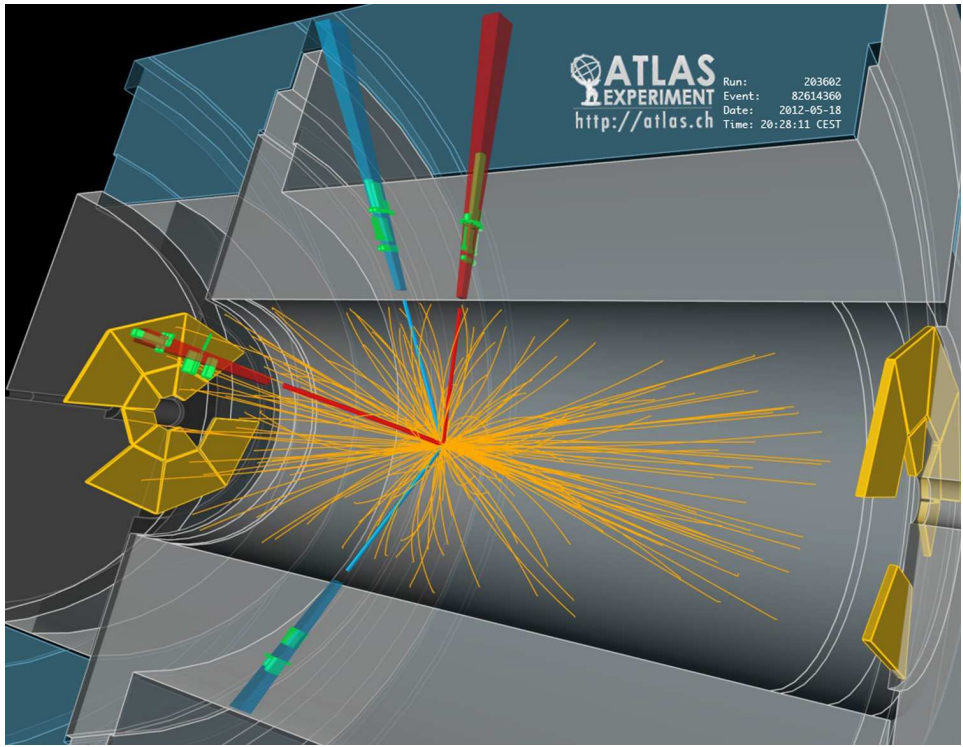}
\caption{An ATLAS candidate for a Higgs decaying into four electrons, indicated by the four long (blue and red) lines, plus other particles. Credit: ATLAS  Collaboration, CERN \label{Fig17-CE0329H.eps}}
\end{figure}

A Higgs is typically created by gluon-gluon fusion when protons collide. The big challenge is to find events that are candidates for the subsequent decay of this Higgs. Two such events are shown in Figs.~\ref{Fig16-CMS-gammas.eps} and \ref{Fig17-CE0329H.eps}. The event from CMS features two photons, and the collision producing four electrons was observed by ATLAS. Of course, each experiment has detected many events of all kinds. Because of the vast number of collisions, it is possible to locate and analyze only a small fraction of the decays of the Higgs-like particle that are undoubtedly occurring -- analogous to needles in an enormous haystack.

\begin{figure}[htbp]
\centering
\includegraphics[bb=0 0 360 270, width=5in]{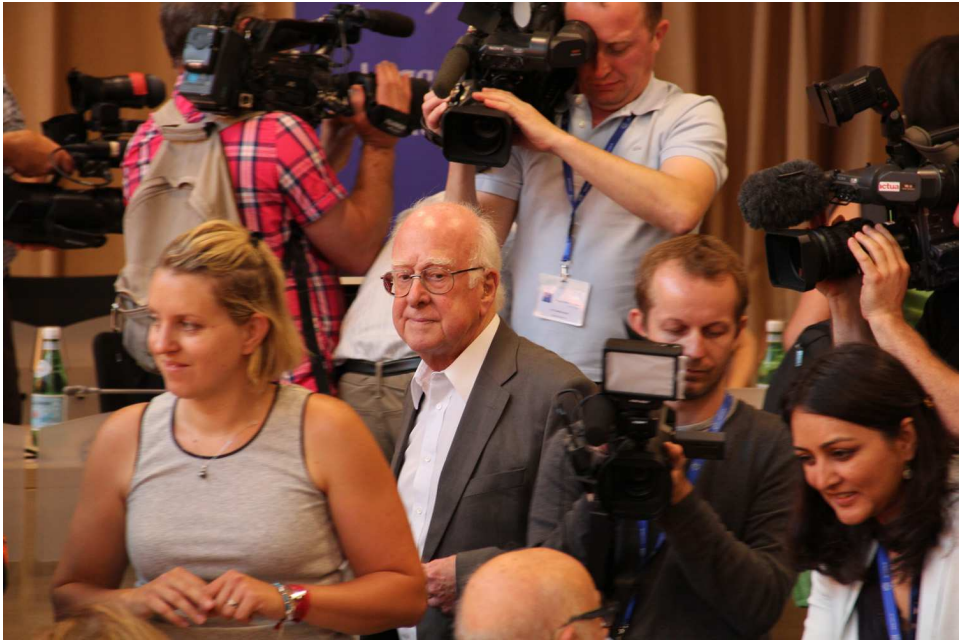}
\caption{Peter Higgs at the announcement of the discovery of a Higgs-like particle. Credit: CERN \label{Fig23-Peter-Higgs.eps}}
\end{figure}

When enough of such events are found that there is a substantial excess over background events (attributable to already understood processes) there is evidence for a Higgs. A very careful statistical analysis is thus required, to back up graphs displaying apparent peaks above background for the various kinds of Higgs decays. When the ``p-value'' reached the 5.0 $\sigma$ level, or about one chance in 3.5 million that statistical fluctuations had mimicked a true discovery, the announcement of the discovery was made, with Peter Higgs present, as can be seen in Fig.~\ref{Fig23-Peter-Higgs.eps}. The later level of certainty was a substantially stronger 6.9 $\sigma$.

\section{\label{sec:sec5a} The unity of physics, from the highest energies to the Standard Model to the lowest temperatures}

It is remarkable that the same laws of physics are valid over billions of light years and down to a small fraction of the diameter of a proton. What is perhaps even more remarkable is that the same basic principles recur again and again in very different regimes. Above it was seen that (1) the masses of SM particles, (2) the masses of neutrinos, and (3) the masses of susy partners are all thought to result from condensation of bosonic fields. This kind of phenomenon has been a recurring theme in condensed matter physics for many decades, and in atomic physics more recently. In all three areas there are now many examples. 

The physics community has gotten used to this prevailing theme, but it is still remarkable that the short range of the weak nuclear force (which results from the large masses of the virtual W bosons that mediate this interaction) has the same basic origin as the Meissner effect in a superconductor, where the photon effectively acquires a mass from the condensate, causing the magnetic field $\boldsymbol{B}$ to be expelled. This idea originates with the 1935 paper of Fritz and Heinz London, who postulated an equation that was later simplified by Fritz London:
\begin{eqnarray}
\mathbf{j}_{s} = - \frac{n_se^2}{mc}\mathbf{A} 
\label{eq4.1}
\end{eqnarray}
where $\mathbf{j}_{s}$ is the current density of the superconducting condensate, $n_s$ is its particle density, $e$ is the electric charge, $m$ is a particle mass, $c$ is the speed of light, and $\mathbf{A} $ is the vector potential of electromagnetism. Then diamagnetic currents are set up which oppose the applied field and allow it to penetrate only a small distance -- the London penetration depth. 
\begin{figure}[htbp]
\centering
\includegraphics[bb=0 0 300 500, width=3in]{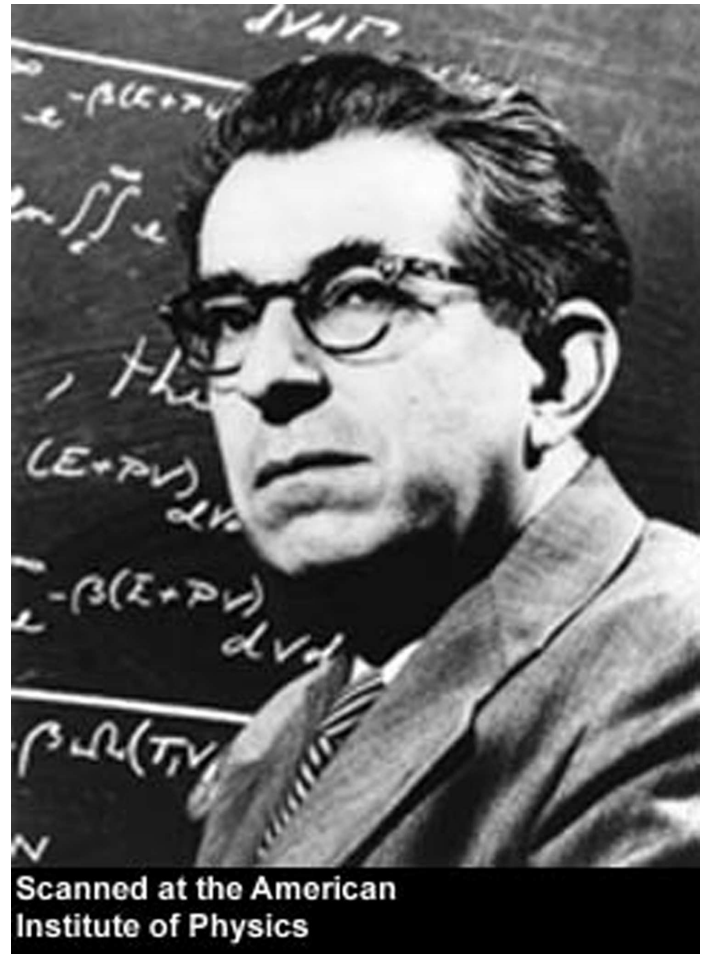}
\caption{Fritz London who, with his brother Heinz, explained the Meissner effect by effectively giving mass to the photon inside a superconductor. Credit: AIP Emilio Segre Visual Archives, Physics Today Collection \label{Fig18-fritz_london.eps}}
\end{figure}

\begin{figure}[htbp]
\centering
\includegraphics[bb=0 0 360 260, width=5in]{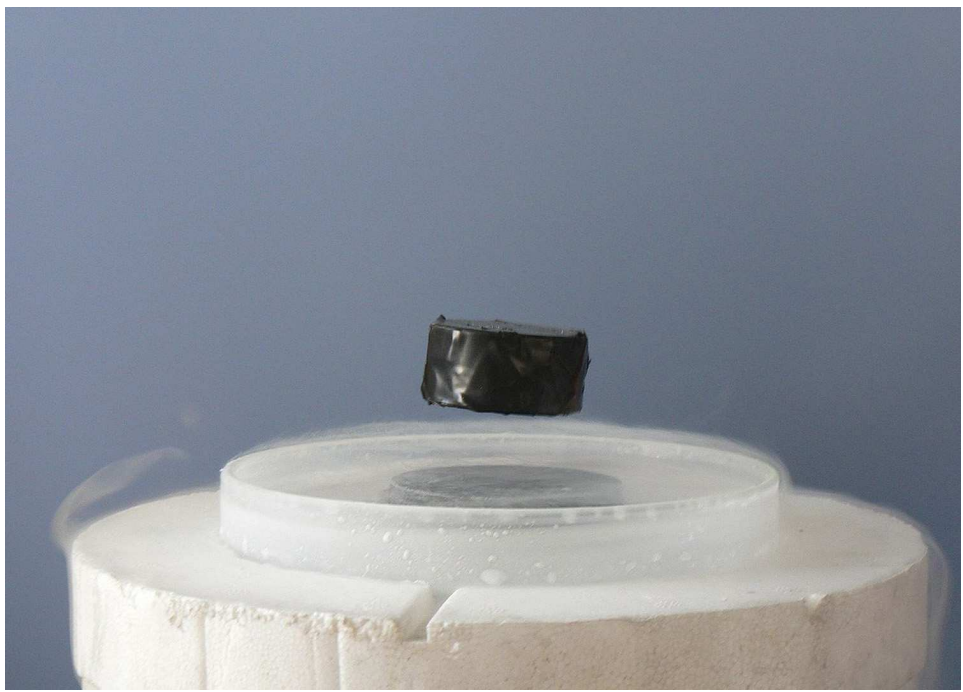}
\caption{A magnet floats above a superconductor, repelled as a result of the Meissner effect. Credit: Credit: (c) Mai-Linh Doan. CC BY-SA \label{Fig19-Meissner.eps}}
\end{figure}

A beautiful demonstration is the levitation of a magnet by a superconductor shown in Fig.~\ref{Fig19-Meissner.eps}.
A modern interpretation of (\ref{eq4.1}) is that the photon is effectively given a mass inside the superconductor. 

In this and the following two paragraphs we momentarily revert to a more technical discussion, involving the idea of gauge invariance that was described in the later part of Section \ref{sec:sec2}. Eq. (\ref{eq4.1}) \textit{effectively} breaks gauge invariance, because a gauge transformation changes $\mathbf{A} $ and not $\mathbf{j}_{s}$.  But gauge invariance turns out to still hold if the ``vacuum'' of the superconducting condensate is included.

There are other examples of this basic concept. In a ferromagnet like iron, rotational symmetry is effectively broken because the magnetic dipole moment points in a particular direction. But overall rotational invariance still holds if the ``vacuum'' of the material is also rotated. 

\begin{figure}[htbp]
\centering
\includegraphics[bb=0 0 360 270, width=5in]{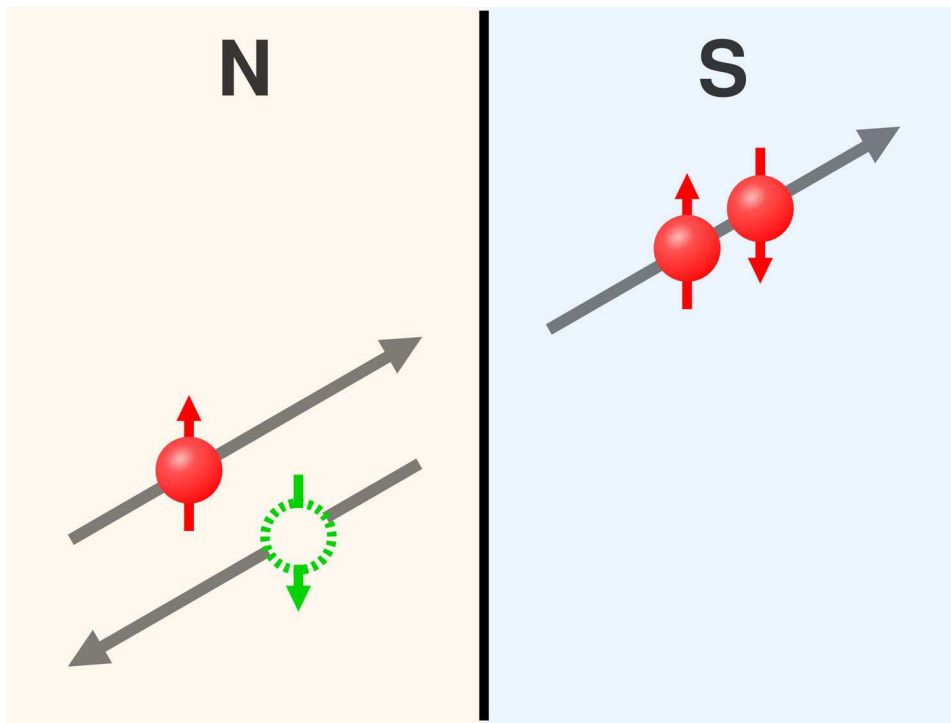}
\caption{Andreev reflection: A negatively charged electron in a normal conductor, labeled N, reflects off a superconductor, labeled S. But it is  reflected as a positively charged hole (with opposite spin and momentum), so that charge conservation is violated if one includes only these particles. But if the condensate of electron pairs in the superconductor is included, charge is conserved. This is a general principle: When fundamental symmetries, and the conservation laws which result from these symmetries, are effectively broken in high energy physics or other areas of physics, they are not truly broken if one includes the ground state, which in the universe is the vacuum. Credit: Ilmari Karonen (for Wikimedia).
\label{Fig20-Andreev.eps}}
\end{figure}

Gauge invariance implies conservation of electric charge, according to Noether's theorem. A wonderful and nontrivial example in condensed matter physics is Andreev reflection off a superconductor, depicted in Fig.~\ref{Fig20-Andreev.eps}. In the effective theory, which omits the condensate or ``vacuum'', charge is not conserved. But in the more fundamental picture the condensate absorbs the missing charge.

\begin{figure}[htbp]
\centering
\includegraphics[bb=0 0 360 250, width=5in]{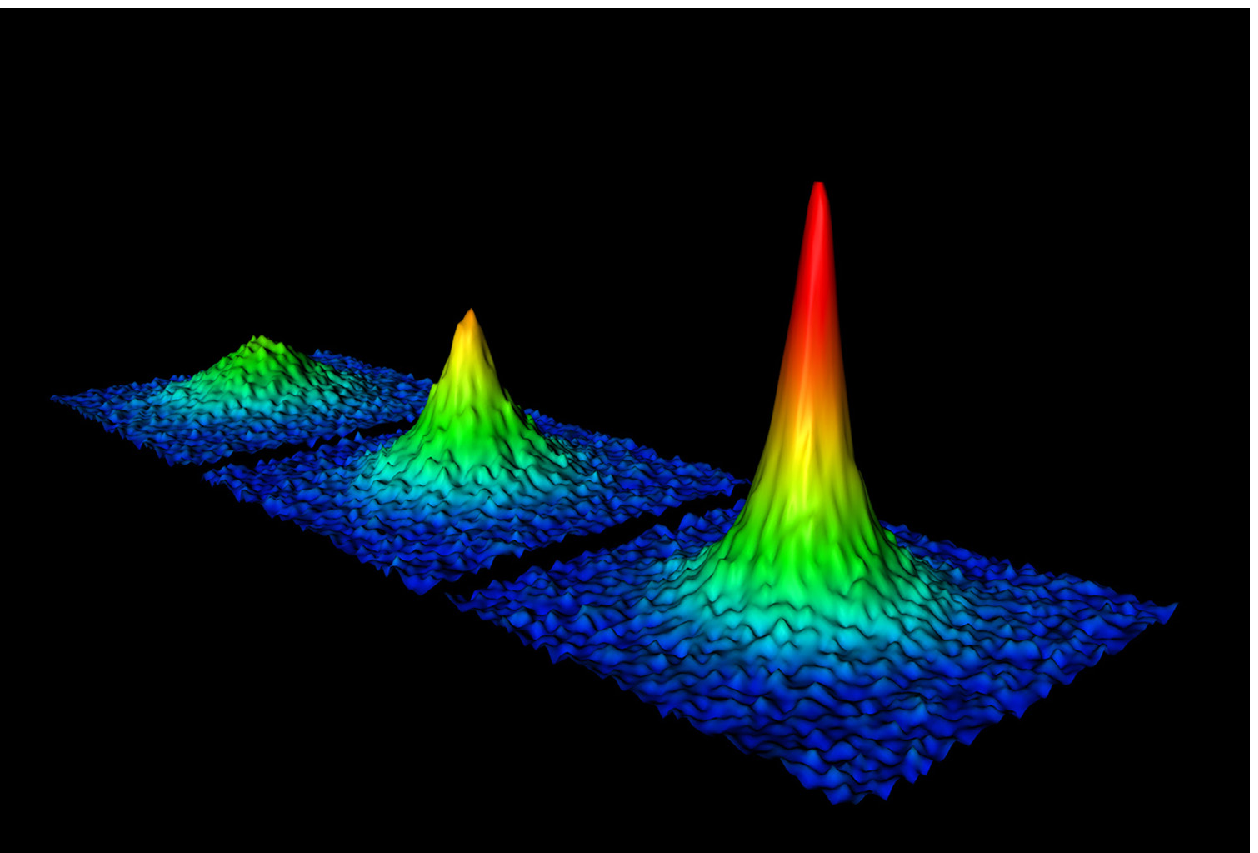}
\caption{Formation of a condensate of pairs of fermionic $^{40}$K atoms as their attraction is strengthened by tuning of a magnetic field. Credit: Markus Greiner, University of Colorado, Boulder. 
\label{Fig22-BEC.eps}}
\end{figure}

Fig.~\ref{Fig22-BEC.eps} illustrates the rather amazing and new ability of laboratories studying bosonic condensates of atoms to cool them to ultralow temperatures (down to half a nanokelvin in one experiment) and then control them with magnetic fields. Here a pair of fermionic atoms effectively behaves as a boson.

\section{\label{sec:sec6} Conclusion}

We live in a universe that is still filled with mysteries, and, in addition to its role as the origin of particle masses, the Higgs is linked to all of them in various ways. It seems to point to susy, as the most natural agent to protect its mass from being increased by many orders of magnitude. The basic Higgs mechanism spills over into the potential origin of masses for neutrinos, and to the condensation of Higgs-like fields in a grand unified theory of forces and particles. It also spills over into the explanation of how supersymmetry is broken and superpartners acquire masses. It is relevant to how inflation might have occurred in the very early universe, via condensation of a Higgs-like inflaton field. 

\begin{figure}[t]
\centering
\includegraphics[bb=0 0 360 300, width=5in]{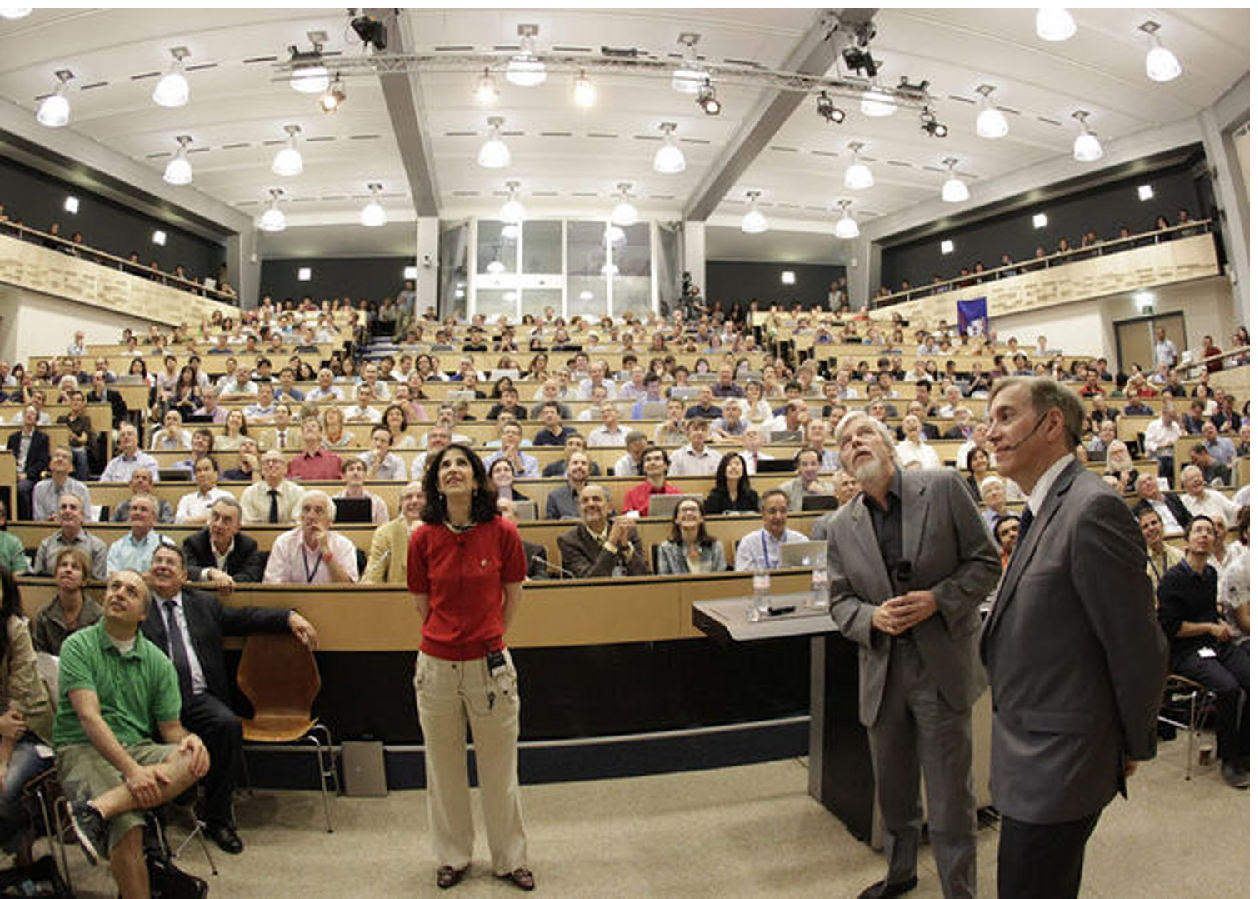}
\caption{The discovery of a new kind of particle is announced, on July 4, 2012, by the spokespersons of the ATLAS and CMS collaborations and the Director General of CERN. Credit: CERN \label{Fig24-discovery.eps}}
\end{figure}

The discovery of a Higgs-like particle was a giant step in the history of science, justifying the 2013 Nobel Prize to Fran\c{c}ois Englert and Peter Higgs~\cite{press}, who perhaps represent all the high-energy (and condensed-matter) theorists who made major contributions. The 6000+ CMS and ATLAS experimentalists, young and senior, and many others probing complementary fundamental issues, still have further discoveries to make, with more moments to come like that recorded in Fig.~\ref{Fig24-discovery.eps}, when the magnitude of this historic discovery was first fully appreciated.

\section*{Acknowledgements}

I am indebted to Alexei Safonov, Bhaskar Dutta, Stephen Parke, Alex Nahmad-Rohen, and Suzanne Lidstr\"{o}m for many helpful suggestions.

\end{document}